\documentclass[12pt]{article}

\usepackage{amsmath,amsfonts,amssymb}
\usepackage{graphicx}
\usepackage{psfrag}
\usepackage{enumerate}
\usepackage{color}

\makeatletter \@addtoreset{equation}{section}

\makeatletter\renewcommand\section{\@startsection {section}{1}{\z@}%
                                   {-3.5ex \@plus -1ex \@minus -.2ex}%nn
                                   {2.3ex \@plus.2ex}%
                                   {\normalfont\large\bfseries}}
\renewcommand\subsection{\@startsection{subsection}{2}{\z@}%
                                     {-3.25ex\@plus -1ex \@minus -.2ex}%
                                     {1.5ex \@plus .2ex}%
                                     {\normalfont\bfseries}}

\newcommand{\be}{\begin{equation}}
\newcommand{\ee}{\end{equation}}
\newcommand{\beq}{\begin{eqnarray}}
\newcommand{\eeq}{\end{eqnarray}}

\def\[{\left [}
\def\]{\right ]}
\def\({\left (}
\def\){\right )}

\def\r2{\sqrt{2}}

\def\t{\tau}

\newcommand{\bbibitem}[1]{\bibitem{#1}\marginpar{#1}}

\newcommand{\figref}[1]{Fig. \ref{#1}}
\newcommand{\secref}[1]{Sec. \ref{#1}}
\newcommand{\trm}[1]{\textrm{#1}}

\def\Label#1{\label{#1}%
  \smash{\hbox to0pt{\raise1ex\hbox{\tiny[#1]}\hss}}}
\def\noLabels{\let\Label=\label}
\def\nobbibitem{\let\bbibitem=\bibitem}

\begin{document}
\noLabels % uncomment for final production
\nobbibitem % uncomment for final production

\begin{titlepage}

%\begin{flushright}%\vspace{-2cm}
%{\small
%UPR-1154-T  \\ %\vspace{-0.35cm}
%LBNL-60486 \\
%hep-th/0606118}%\\
%\end{flushright}
%\vspace{12 mm}

\vfil\
%vfil

\begin{center}

{\Large \bf Bubble, Bubble, Flow and Hubble: \\[.25cm] \large Large Scale Galaxy Flow from Cosmological Bubble Collisions}

\vspace{3mm}

Klaus Larjo$^{a,}$\footnote{email: larjo@phas.ubc.ca} and Thomas S.
Levi $^{a,b,}$\footnote{email: tslevi@phas.ubc.ca}
\\

\vspace{5mm}

\bigskip\medskip
\centerline{$^a$\it Department of Physics and Astronomy}
\smallskip\centerline{\it University of British Columbia}
\smallskip\centerline{\it Vancouver, BC V6T 1Z1, Canada}
\bigskip \centerline{$^b$\it Center for Cosmology and Particle Physics}
\smallskip\centerline{\it Department of Physics, New York University}
\smallskip\centerline{\it 4 Washington Place, New York, NY 10003.}

\vfil

\end{center}
\setcounter{footnote}{0}
%%%%%%%%%%%%%%%%%%%%%%%%%%%%%%%%%%%%%%%%%%%%%%%%%%%%%%%%%%%%%%%%%%%%%%%%%%%%%%%%%%%%%%%
\begin{abstract}
\noindent
We study large scale structure in the cosmology of Coleman-de Luccia bubble collisions. Within a set of controlled approximations we calculate the effects on galaxy motion seen from inside a bubble which has undergone such a collision. We find that generically bubble collisions lead to a coherent bulk flow of galaxies on some part of our sky, the details of which depend on the initial conditions of the collision and redshift to the galaxy in question. With other parameters held fixed the effects weaken as the amount of inflation inside our bubble grows, but can produce measurable flows past the number of efolds required to solve the flatness and horizon problems.

\end{abstract}
%%%%%%%%%%%%%%%%%%%%%%%%%%%%%%%%%%%%%%%%%%%%%%%%%%%%%%%%%%%%%%%%%%%%%%%%%%%%%%%%%%%%%%%%%
\vspace{0.5in}

\end{titlepage}
\renewcommand{\baselinestretch}{1.05}  %Line spacing
%%%%%%%%%%%%%%%%%%%%%%%%%%%%%%%%%%%%%%%%%%%%%%%%%%%%%%%%%%%%%%%%%%%%%%%%%%%%%%%%
%%%%%%%%%%%%%%%%%%%%%%%%%%%%%%%%%%%%%%%%%%%%%%%%%%%%%%%%%%%%%%%%%%%%%%%%%%%%%%%%%%%%%%%%%%%
\tableofcontents

\newpage

\section{Introduction}
One of the greatest challenges for string theory is to find a testable prediction which can differentiate it from other potential theories of quantum gravity and high energy physics. Unfortunately, the typical energy scales at which string theoretic effects become important is likely far out of reach for any earth-based particle accelerator. Cosmology though, provides another opportunity. The early universe was arbitrarily hot and dense and signatures of the physics at that time remain in the large-scale structure of the universe today.

It has become increasingly clear in recent years that one of the characteristic features of string theory is the ``landscape'' model of cosmology \cite{discretum,kklt,lennylandscape}.  There are many ways to generate realistic four-dimensional universes in string theory via flux compactifications. Various parameters specifying the compactification become scalar fields in our four-dimensional universe. The landscape is a potential energy function of the would-be moduli scalar fields, which possesses many distinct minima.  These minima are meta-stable and can decay via bubble nucleation-type tunneling or quantum thermal processes.  In the early universe, one expects all the minima to be populated as the temperature decreases and the fields begin to settle.  The minima with the highest values of potential energy then begin to inflate, and are expected to rapidly dominate the volume.  After a few string times the universe will be almost completely full of rapidly inflating regions, and as time continues to pass the metastable field values in these regions will begin to decay by bubble nucleation.  Since we live in a region with very small vacuum energy, we expect that our observable universe is contained in such a bubble.

The most promising way to test this idea is to look at large scale cosmological structure.  If we model the string theory landscape as a scalar field coupled to gravity with a potential with several minima, universes inside a bubble which formed via a tunneling instanton have certain characteristic features:  they are open, they are curvature dominated at the ``big bang" (rendering it non-singular), and they have a characteristic spectrum of density perturbations generated during the inflationary period that takes place inside them  \cite{cdl,gott1,gott2,Garriga:1998he,lindeopen,Freivogel:2005vv,batra}.

Moreover, because the bubble forms inside a metastable region, other bubbles will appear in the space around it.  If these bubbles appear close enough to ours, they will collide with it \cite{oldguth,ggv,Aguirre:2007an}.  The two bubbles then ``stick" together, separated by a domain wall, and the force of the collision releases a pulse of energy which propagates through each of the pair.  This process (as seen from inside our bubble) makes the universe anisotropic, affects the cosmic microwave background and the formation of structure.  A collision between two bubbles produces a spacetime which can be solved for exactly \cite{ben,wwc,Aguirre:2007wm,wwc2}.

In \cite{wwc,wwc2}, one of these authors began the study of this process and its effects. In these papers we principally studied the effects of such a collision on the cosmic microwave background (CMB).  We found that bubble collisions naturally create hemispherical power asymmetries spread through the lower CMB multipoles and/or produce relatively small cold or hot spots, depending on the time of the collision (and hence the size of the disk), and can lead to measurable non-Gaussianities.  The qualitative features, however, are always the same:  a collision with a single other bubble produces anisotropies that depend only on one angle---the angle to the vector pointing from our location to the center of the other bubble---and which are generically monotonic in the angle.

While the CMB temperature map is a good probe of these effects, it is by no means the only one. In this paper, we begin to look
for other measurable effects by studying how a single bubble collision affects large scale structure and
galaxy motion. We find that collisions generically lead to a large scale coherent flow of galaxies in our sky. The size of the
affected disk of galaxies, and the magnitude of the effect are controlled by similar order parameters for the CMB effect, as well
as the redshifts of the galaxies in question. Generically, the angular size of the affected disk on the sky for large scale
structure is always larger than the corresponding disk for the CMB hot/cold spots. It is very interesting to consider our results
in light of recent observational measurements of bulk galaxy flows dubbed ``dark flow''
\cite{darkflow,Kashlinsky:2008us,Watkins:2008hf}.

Our model consists of a collision between our bubble -- which we treat as a thin-wall Coleman-de Luccia bubble which undergoes a
period of inflation after it formed (which we model as a de Sitter phase) followed by a period of radiation domination -- and
another different bubble, the details of which are largely unimportant.  We assume that the tension of the domain wall between
the bubbles and the vacuum energy in the other bubble are such that the domain wall accelerates away from us.  Such collisions
can produce observable effects which are not in conflict with current data.  We will also assume that the inflaton is the field
that underwent the initial tunneling transition, so that the reheating surface is determined by the evolution of that field and
can be affected by the collision with the other bubble through its non-linear equation of motion, and that no other scalar fields
are relevant.

After some number of efolds of inflation, our bubble will reheat and become radiation dominated. The subsequent expansion and the
geometry of the reheating surface itself will be modified by the presence of the domain wall. In \cite{wwc2}, it was shown how
the surface was modified. This modified surface will also backreact on the spacetime because the various fluids created at
reheating will no longer be classically comoving. In \cite{wwc2} this backreaction was ignored and  it was assumed the universe
could be approximated as standard radiation dominated Friedmann-Robertson-Walker (FRW) universe for the subsequent evolution. We
will show here that this approximation is valid for the CMB photons, but not for the matter fluid that will form
the galaxies we are interested in here. To find the evolution of galaxies, some measure of this backreaction must be taken into
account, and we make a first approximate attempt to do so here.

There are a number of anomalies in current cosmological data, including a cold spot \cite{coldspot1, coldspot2, coldspot3, coldspot4, coldspot5}, hemispherical power asymmetry \cite{Eriksen:2007pc}, non-Gaussianities (see {\it e.g.} \cite{nongauss}),  the ``axis of evil" \cite{Land:2005ad}, and the tentative possibility of a ``dark flow" \cite{darkflow,Kashlinsky:2008us,Watkins:2008hf}.  While the significance of these effects is unclear, there are few if any first-principle models which can account for them.  Our results here, combined with \cite{wwc,wwc2} indicate that bubble collisions can naturally create hemispherical power asymmetries spread through the lower CMB multipoles, relatively small cold or hot spots, measurable non-Gaussianities and a possible macroscopic, coherent bulk flow of galaxies. While it is not yet clear that a single collision can produce all of these effects in agreement with observations, it is nevertheless interesting that bubble collisions can at least qualitatively produce many of these effects.

\section{Basic Setup} \label{sec-setup}
In this section we will set up the basic scenario for the bubble collision and subsequent evolution we wish to study. The details
of the dynamics of general collisions were studied in \cite{wwc}, and some effects on the CMB were studied in \cite{wwc2}. We
refer the reader there for a more complete treatment. We will focus on a collision scenario where our bubble is dominated by
positive vacuum energy at formation, so that it undergoes a period of inflation. We will further assume the pressures are such
that domain wall that forms between the two bubbles is accelerating away from the center of our bubble at late times. This second
assumption is valid for collisions between two dS bubbles if ours has a smaller cosmological constant.\footnote{It also occurs
when our bubble has a larger cosmological constant or in collisions with AdS bubbles when certain conditions involving the
tension of the wall are satisfied \cite{ben,wwc,Aguirre:2007wm,Cvetic:1996vr}.} While more general scenarios are possible, these
assumptions lead to an evolution that is consistent with current observations.

\begin{figure}[t]
\begin{center}
\includegraphics[width=0.7\textwidth]{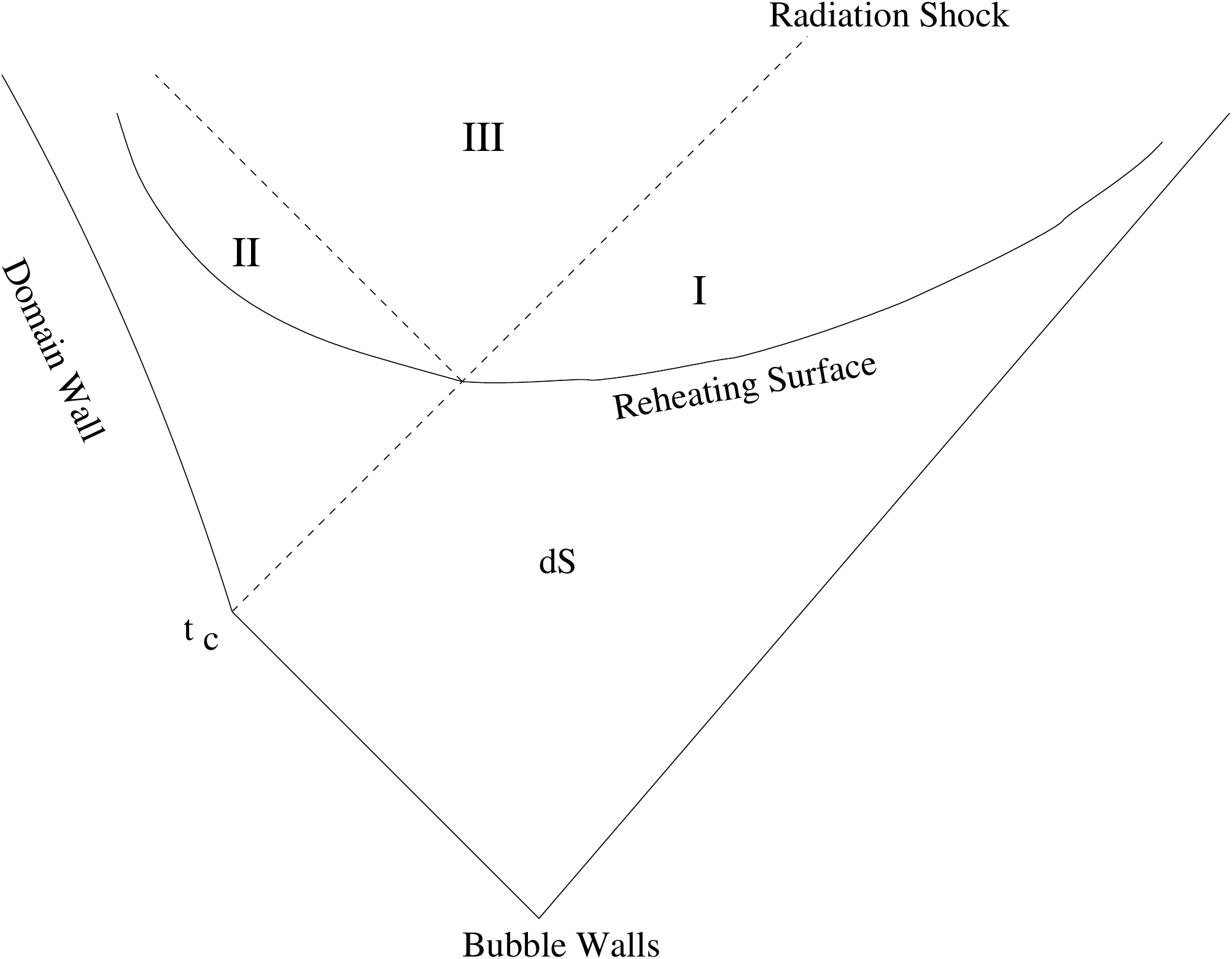}
\end{center}
\caption{\label{fig-setup} Sketch of the causal structure of the spacetime in the bubble collision scenario. We define $t_c$ and regions I, II and III of our bubble in \secref{sec-rehflow}}
\end{figure}

We study the motion of galaxies and large scale structure in this scenario. A full treatment would require solving the
full cosmological perturbation theory and evolution taking into account the macroscopic anisotropy caused by the collision. This
treatment is outside the realm of the current paper. Instead, we will make several assumptions so that an approximate analytic
treatment is possible. Our basic scenario unfolds as follows: our bubble undergoes a period of inflation sufficient to solve the
horizon, flatness and relic problems  in line with current observations ($N \gtrsim 65$ e-folds) after which the spacetime
reheats. After reheating the universe can be approximately described by a radiation dominated (RD) Friedmann-Robertson-Walker
(FRW) universe. We say approximately, because as shown in \cite{wwc2} the reheating surface is modified by the bubble collision
and formation of the domain wall. In short, the domain wall causes the reheating surface to ``bend'' towards or away from the
domain wall depending on the details of the collision. See \figref{fig-setup} and \cite{wwc2} for more details. Because the universe
now reheats along a deformed, non-comoving surface the spacetime is not truly RD-FRW, as some of the symmetry of this geometry is
broken. In \cite{wwc2} we still approximated the subsequent evolution as taking place in RD-FRW. For CMB photons, this is a
reasonable approximation (as will be shown later).  However, for the matter fluid and galaxies,
this approximation is no longer justified and some measure of the backreaction and modified geometry must be taken into account.
We will attempt to do so here. In reality, the spacetime will also undergo a transition to matter domination at some later time
after last scattering. We ignore this latter transition in this paper, as we believe the essence of the galaxies motion is
captured in our setup. It would nevertheless be interesting to study this in more detail and see if our assumptions hold.

To find the approximate motion of galaxies, we assume that the initial conditions of the matter fluid are that it is moving
orthogonal to the reheating surface. We will then attempt to both capture the backreaction of this fluid on the spacetime
(thereby modifying our spacetime from RD-FRW) and track the motion of galaxies through this new spacetime by computing the
timelike geodesics for each galaxy. We are thus making several assumptions, some of which we outline here:

\begin{itemize}
\item{We compute the backreaction by assuming that the local geometry around any point on the reheating surface is RD-FRW. However, each patch will not necessarily be comoving with respect to other patches and so a global description as RD-FRW is impossible.}
\item{We do not account for further evolution of the backreaction effect, i.e. self interaction and gravitation of the radiation fluid and its subsequent effect on the geometry. We expect that these will have a negligible effect on the galaxy motion we are interested in here.}
\item{We ignore self-interaction of the matter fluid/galaxies and instead compute freestreaming geodesics in the modified geometry.}

\end{itemize}

\subsection{Review of geometries}
In this section we will give a brief review of the geometries and establish our notation conventions, for more details as well as
a full description of the causal structure the reader is directed to \cite{wwc,wwc2}.

A single Coleman-de Luccia bubble in four dimensions has an $SO(3,1)$ invariance, inherited from the $SO(4)$ of the Euclidean
instanton \cite{cdl}. When another bubble collides with ours a preferred direction is picked out, breaking the symmetry down to
$SO(2,1)$. As a result one can choose coordinates in such a way that the metric takes the form of a warped product between
two-dimensional hyperboloids ($H_2$) and a 2D space.

Under the assumption that all the energy in the spacetime remains confined to thin shells or is in the form of vacuum energy, one can find the solution describing the collision by patching together general solutions to Einstein's equations in vacuum plus cosmological constant with this symmetry \cite{wwc,ben,Hawking:1982ga,Wu:1984eda}. The overall result is that the spacetime inside our bubble is divided into two regions--one that is outside the future lightcone of the collision and can be described using an $SO(3,1)$ invariant metric (a standard open FRW cosmology), and one which is inside, is affected by backreaction, and is most conveniently described in the general case using 2D hyperboloids. However, we will find that at the late times important in our analysis, it becomes possible to once again use 3D hyperboloids and in our approximation 3D flat space (that is the geometry will be approximately {\it locally} flat FRW).

\paragraph{de Sitter space:}
The general $SO(2,1)$ invariant spacetime with a positive cosmological constant is given by \cite{wwc}
\beq%
ds^2 = {-dt^2 \over g(t)} +g(t) dx^2 +t^2 dH_2 ^2,%
\eeq%
where
\beq%
g(t) = 1 +{t^2 \over \ell^2} -{t_0 \over t} ,%
\eeq%
and%
\beq%
dH_2^2 = d\rho^2 + \sinh^2 \rho d\varphi^2 %
\eeq%
is the metric on the unit $H_2$ and we identify $x \simeq x+ \pi \ell$. We are most interested in observers far from the domain wall, as these see cosmologies most like ours.
In that case it was shown in \cite{wwc2} that for the collision to be within our past lightcone we require $t_0 \ll t$ at cosmologically relevant times.
Therefore, throughout the rest of the paper we approximate $t_0 \approx 0$ giving $g(t) \approx 1+ t^2/ \ell^2$.\footnote{It is important to note that $t_0/t$ may {\it not}
be small at early times when some fluctuations left the horizon. It would be interesting to investigate what effect this has on the CMB and other observables.} This is a metric for
pure dS space (albeit in unconventional coordinates) with radius of curvature $\ell$. Hereafter, we set $\ell=1$; it can easily be restored with dimensional analysis. These coordinates,
which we denote the $H_2$ coordinates, are related to the $SO(3,1)$ or $H_3$ invariant ones by%
\begin{eqnarray}
\cosh \tau & = & \sqrt{1+t^2} \cos (x-x_0), \nonumber \\
\sinh \tau \, \sinh \xi \, \cos \theta & = & \sqrt{1+t^2} \sin (x-x_0), \Label{H2-H3} \\
\sinh \tau \, \cosh \xi & = & t \cosh \rho, \nonumber  \\
\sinh \tau \, \sinh \xi \, \sin \theta & = & t \sinh \rho, \nonumber \\
\varphi &=& \varphi,
\end{eqnarray}
where the free parameter $x_0$ is used to fix the location of the origin of $H_3$. In these coordinates the metric becomes
\beq%
ds_{dS}^2 & = & - d\tau^2 + \sinh^2 \tau \, dH_3^2, \quad \textrm{with} \Label{FRW} \\
dH_3^2 & = & d\xi^2 + \sinh^2 \xi \, (d\theta^2 + \sin^2 \theta \, d\varphi^2) \, .\nonumber
\eeq%
It was also shown in \cite{wwc2} that an observer can only see an exponentially small part of the reheating surface in $\xi$, that is the $\Delta \xi$ an observer
has access to is very small (we will make this more precise shortly). Thus, if we choose $x_0$ appropriately we have $\xi \ll 1$ for any observer. Throughout we make this approximation and we thus find%
\beq%
dH_3^2 & = & d\xi^2 + \sinh^2 \xi \, (d\theta^2 + \sin^2 \theta \, d\varphi^2)  \approx  d\xi^2 + \xi^2 \, (d\theta^2 + \sin^2 \theta \, d\varphi^2) = ds^2_{\mathbb{R}_3} .
\eeq%

\paragraph{Radiation domination:}  We will also make use of the RD-FRW metric given by%
\begin{eqnarray}
ds^2_{\textrm{RD}} & = & -d\tau^2 + a^2(\tau) \, dH_3^2 \approx -d\tau^2 + a^2(\tau) \, ds_{\mathbb{R}_3}^2,  \Label{eq:RD} \\
a(\tau) & = & \sqrt{C(\tau + \frac{1}{2})},
\end{eqnarray}
with $C$ a constant.\footnote{We find it convenient to break slightly with the conventions of \cite{wwc2} by translating
$\tau_{\textrm{RD}}$ by $\frac{1}{2}$ above; this will simplify subsequent expressions. }

At reheating we need to match both the scale factor $a$ and Hubble constant $H$ of the two metrics \cite{wwc2}. This gives%
\beq%
\tau^{\textrm{RD}}_r & = & \frac{1}{2} \left( \tanh \tau^{dS}_r - 1\right), \Label{rehtimes}\\
C & = & 2 \cosh \tau^{dS}_r \, \sinh \tau^{dS}_r = 2 a_r^2,
\eeq%
where $a_r \equiv a(\tau^{RD}_r)$ is the scale factor at reheating. Note that (\ref{rehtimes}) is a gluing relation joining
the two spacetimes; it should not be interpreted as a coordinate transformation. We also note that we are able to use this
solution only because $t_0/t \ll 1$ for the cosmologically relevant times. If this were not the case, the solution would be
more complicated. Due to the reduced symmetry, the general solution for RD with $SO(2,1)$ invariance is not known.\footnote{The
problem is roughly equivalent to finding the smooth metric describing a Schwazschild black hole embedded in an asymptotically FRW
spacetime.}

\paragraph{Parameters and expansions:}
It will be useful here to define our various parameters and orders of expansion. The number of efolds of inflation is $N \equiv \tau^{dS}_r$. After reheating the temperature redshifts as $1/a$ so we can define%
\beq%
e^{N_*} \equiv T_r / T^0_n = a_n /a_r ,%
\eeq%
where the $n$ subscript refers to the observer's time and this defines $N_*$. In \cite{wwc2} it was shown that an observer can
see a part of the reheating surface approximately $\Delta \xi \approx 4 e^{N_*-N}$ in size. As $\Delta \xi \sim 1$ is one curvature
radius of the $H_3$,  $N_*$ is roughly the number of efolds required to solve the flatness and horizon problems. To agree
with current observations we require $N, N_* \gg 1$ and $N_* < N$. This justifies the previous assumption that $\Delta \xi \ll
1$. In our computations we frequently drop subleading terms that are suppressed by $e^{-N}$ or $e^{-N_*}$, and expand all quantities
as power series in
\be%
\epsilon \equiv e^{N_*-N} \ll 1 .
\ee%
We will work to first subleading order in $\epsilon$ throughout this paper.

\section{Reheating and matter flow} \label{sec-rehflow}
In this section we review and discuss the geometry of the reheating surface, and its effect on the post-reheating spacetime. A
full treatment of the reheating surface is given in \cite{wwc2} and for details the reader is directed there. Here we will present a brief review. The
basic idea is as discussed in \secref{sec-setup}: the collision and domain wall modify the reheating surface and break the
$SO(3,1)$ isotropy to $SO(2,1)$. We define the reheating surface as a surface of constant  energy density for the inflaton field
$\phi$. During inflation these surfaces are also surfaces of constant value for the inflation field. This agrees with the usual
slow-roll conditions so long as we are far enough away from the domain wall. Near the wall, the inflaton develops large spatial
gradients and the usual slow-roll treatment is not enough. In this paper, we will focus on the region where $\phi=
\textrm{constant}$ is enough to define the reheating surface.\footnote{Work on the strongly coupled region is in progress
\cite{endinf}.}

\subsection{The scalar field and reheating surface}
In \cite{wwc2} an approximate solution for the scalar field and reheating surface was presented.\footnote{A full treatment
requires solving the non-linear coupled scalar field and gravity system for the collision spacetime. Some work in a special case
was presented in \cite{Aguirre:2008wy}.} This computation involves several steps. We need to solve the scalar field equation with
a linear potential, $V(\phi)=\mu \phi$, which is a good approximation during slow roll inflation.

The remaining inputs are the boundary conditions. We expect the scalar field to have a set of values $\phi_L$ in the other
bubble, some other set of values $\phi_R$ in our bubble and a third set in the metastable vacuum outside both bubbles $\phi_M$.
In the thin wall limit the scalar field jumps from near some other local minimum (either the false vacuum our bubble tunneled
from, or the vacuum inside the bubble it collided with) to a location a bit up the ``hill'' from our local minimum. This
transition occurs all along the domain and bubble walls, so the scalar field is constant (though not necessarily the same
constant) along each of these surfaces. As in \cite{wwc2} we choose without loss of generality $\phi=0$ along the bubble wall and
$\phi=k$ along the domain wall. Note that $k=0$ is a possible boundary condition. The remaining condition on $\phi$ is that it is
continuous across the radiation shell of the collision. At linear order this is enough to give us the solution for $\phi$ and the
reheating surface as a function of $k$, the potential and the details of the collision (such as collision time, acceleration of
the domain wall, etc.). Once again for a full treatment the reader is directed to \cite{wwc2}, where the full solution for  $\phi(t,x)$
is given.

For reasons stated above, we are interested in the reheating surface near the radiation shock, as illustrated in
\figref{fig-setup}. The null line for the radiation shock was found in \cite{wwc2} as
\begin{equation}
x_{\textrm{rad}} = \arctan t - 2 \arctan t_c \approx \frac{\pi}{2} - 2 \arctan t_c, \Label{xrad}
\end{equation}
where $t_c$ is the time of the collision, and we approximated $\arctan t \approx \frac{\pi}{2}$, as reheating happens after
$t\sim e^N \ggg 1$. Expanding the solution for the inflaton $\phi(t,x)$ around $x = x_{\textrm{rad}}$ we find
\begin{equation}
\phi(t,x) \approx - \frac{\mu}{3} \ln \left[ t \cos (x-x_{\textrm{rad}}) \right] -{\mu \over 3} A
(x-x_{\textrm{rad}})\Theta(-(x-x_{\textrm{rad}})) + \mathcal{O}((x-x_{\textrm{rad}})^2), \Label{phi-h2}
\end{equation}
where $\phi_{\textrm{ref}} = - \frac{\mu}{3} \ln \left[ t \cos (x-x_{\textrm{rad}}) \right] $ corresponds to the solution in the
unperturbed case, without a collision. The step function ensures the solution is given by $\phi_{\textrm{ref}}$ in the part of
spacetime causally disconnected from the collision. The coefficient of the linear term is given by
\begin{equation}
A \equiv -\frac{1}{2} \left(  t_c - \frac{1}{t_c} - 4\, \frac{\frac{3k}{\mu} + \ln \left[ \frac{4t_c}{1+t_c^2} \frac{1}{\pi - 2
\arctan t_c} \right]}{\pi - 2 \arctan t_c} \right),
\end{equation}
up to corrections of order $\frac{1}{t} \sim e^{-N}$.  For realistic values of the parameters $(\mu, k, t_c)$, the slope $A$ is
of the order $\mathcal{O}(1)$. In the rest of this paper we will be working to first order in the parameter $A
(x-x_{\textrm{rad}}) \sim \mathcal{O}(e^{N_*-N}) \ll 1$.\footnote{The coefficient of the quadratic term
in (\ref{phi-h2}) is also of order $\mathcal{O}(1)$, and hence the linear approximation is justified.}

It is useful to convert this to $H_3$ coordinates for use in the rest of the paper. Using the coordinate transformation
(\ref{H2-H3}) with the origin of $H_3$ fixed to the radiation shock by setting $x_0=x_{\textrm{rad}}$, we find
\begin{equation}
\phi(\tau,\xi,\theta)  \approx  \phi_{\textrm{ref}}(\tau) - \Theta(-\cos \theta) \left[ {\mu \over 3} A \, \xi \cos \theta  +
\mathcal{O}(\xi^2)\right] -{\mu \over 3} \ln 2, \Label{phi-expand}
\end{equation}
where $\phi_{\textrm{ref}} (\tau) =  -\mu \tau/3$ and we can drop the $\ln 2$ term without loss of generality as it amounts to a constant shift of $\phi$ and has no effect on the reheating surface. Hence in these coordinates  the reheating surface consists of two flat 3D-surfaces meeting at an angle, as indicated in
\figref{fig-as6}.

The form of the scalar field in (\ref{phi-expand}) suggests that it will be convenient to switch to a cartesian system
\begin{equation}
\left\{ \begin{array}{l} \xi = \sqrt{x^2 + y^2}, \\  \theta = \arctan \frac{y}{x}, \end{array} \right. \quad \Leftrightarrow
\quad \left\{ \begin{array}{l} x = \xi \cos \theta, \\ y = \xi \sin \theta, \end{array} \right.
\end{equation}
so that $\phi$, and thus the natural equal time surfaces, only depend on $\tau$ and $x$.\footnote{This $x$ should not be confused with the
$x$ in $H_2$ coordinates; from now on we'll only use $H_3$ coordinates and $x$ will be as defined above.}

\subsection{Matter flow and causal regions}
\begin{figure}[h]
\begin{center}$
\begin{array}{cc}
\includegraphics[width=0.4\textwidth]{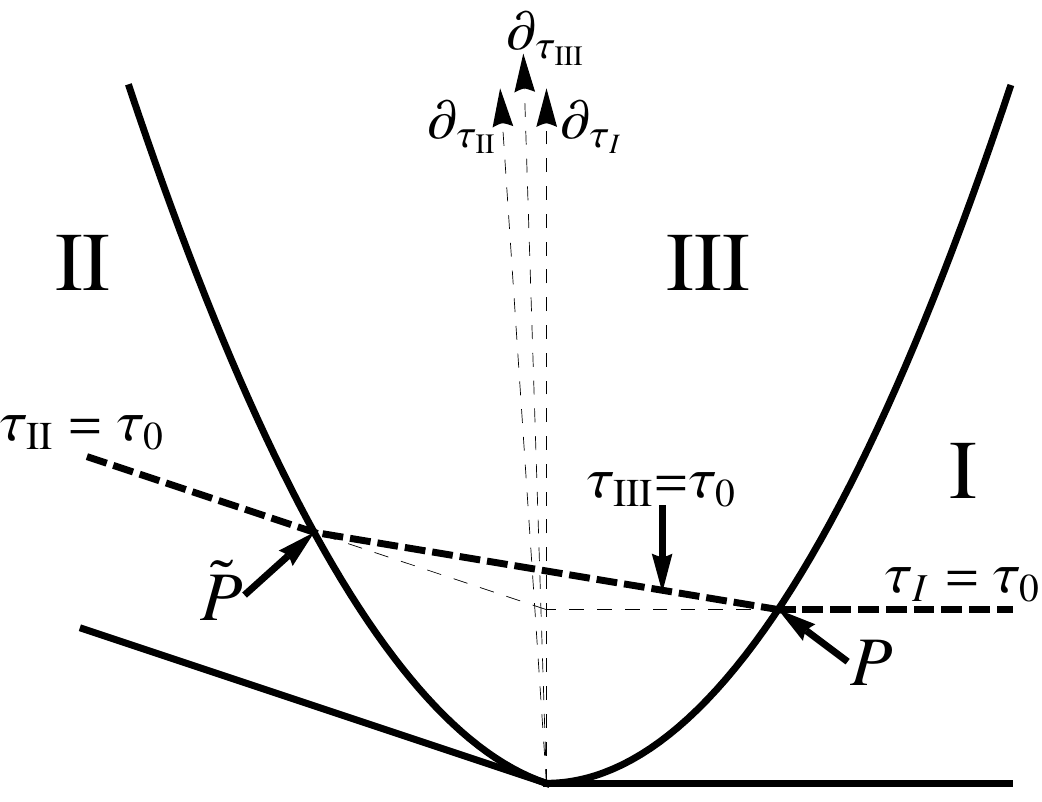} \quad \qquad & 
\includegraphics[width=0.4\textwidth]{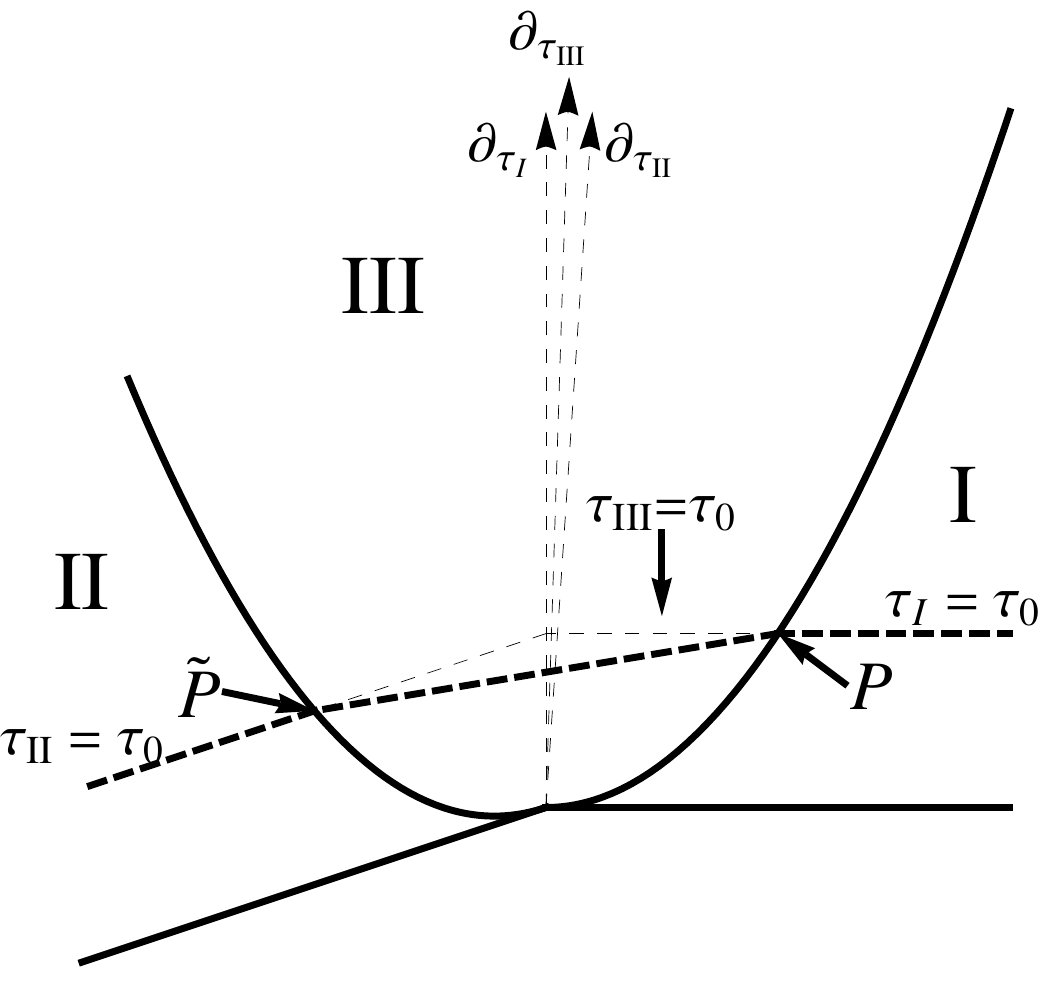}
\end{array}$
\end{center}
\caption{\label{fig-as6} Spacetimes after reheating, showing the reheating surface, an equal time surface $\tau_0$, and the three
time axes. The tilting of the time axes is exaggerated to distinguish between them; in reality they would be almost parallel for
large $a_r$. The figure on the left (right) corresponds to $A>0 \, (<0)$ and would lead to a hot (cold) spot in the CMB.}
\end{figure}

The spacetime after reheating is separated into three causal regions, denoted I, II and III in Figs. \ref{fig-setup} and \ref{fig-as6}.  Observers
in region I are causally disconnected from the bubble collision, and the spacetime in region I is RD-FRW space with time
coordinate $\tau_I \equiv \tau$, as given in (\ref{eq:RD}). Since region II is casually disconnected from region I, observers in region II only see the perturbed part of the reheating surface.
The metric in region II is sourced by the perturbed part of reheating surface only. The symmetry of the reheating
surface then implies that region II is also given by RD-FRW space, but with a time coordinate $\tau_{II}$ that is constant on this part of the surface. One way to view this region is as a RD-FRW space that has been boosted to be comoving with respect to the sloped reheating surface. 

Our first goal will be to seek a relation between the II and I coordinates to find a global description of these regions (in fact, we will find these relations closely resemble a Lorentz boost). In region III, the metric is unknown. A full treatment would backreact the interacting radiation fluid in this region, which is beyond the scope of the current paper. Instead, we will attempt to approximate the backreacted metric in region III by taking a well motivated interpolation between regions I and II. More details on this interpolation will be given in \secref{sec-reg3}

We take the matter fluid to be created at reheating, and moving orthogonal to the reheating surface. Hence, in regions I and II
the matter fluid moves along geodesics defined by $\partial_{\tau_I}$ and $\partial_{\tau_{II}}$ respectively. This assumption is
justified as any peculiar velocity given to the matter field at reheating will quickly redshift away in RD-FRW space, leaving the
motion along $\partial_{\tau_I}$ and $\partial_{\tau_{II}}$ in regions I and II to a good approximation.

\subsection{Relations between the regions and metrics}

\paragraph{Region II:}
From the form of the scalar field (\ref{phi-expand}), and the gluing relation (\ref{rehtimes}), we find that the reheating
surface is given in RD coordinates as
\begin{equation}
\textrm{Reheating surface: }\quad \left\{ \begin{array}{ll} \tau = 0, & \textrm{for } x \ge 0, \quad \textrm{and} \\ \tau + A x =
0, & \textrm{for } x < 0.
\end{array} \right. \Label{reh}
\end{equation}

To define the coordinates in region II we note that $\tau_{II}=0$ on the sloped part of the reheating surface (\ref{reh}), which
together with the space being RD-FRW implies
\begin{equation}
\tau_{II} = \gamma(\tau_I + Ax_I)
\end{equation}
 for some constant $\gamma$. In a coordinate basis we have $\langle d\tau_{II},\partial_{x_{II}} \rangle = 0$, where $\langle , \rangle$ denotes the inner product with respect to the metric, and thus
 \begin{equation}
 \partial_{x_{II}} = \tilde{\gamma}
(A\partial_{\tau_I} -\partial_{x_I})
\end{equation}
 for some $\tilde{\gamma}$. We can fix the constants of proportionality at the intersection of the two parts of the reheating surface, as the metric there is given by RD-FRW in both coordinates, and we have
\begin{eqnarray}
-1 & = & \langle d\tau_{II},d\tau_{II}\rangle = \gamma^2 \langle d\tau_I+A dx_I,d\tau_I+Ad x_I\rangle = \gamma^2 (-1 +
\frac{A^2}{a_r^2}), \nonumber \\
a_r^2 & = & \langle \partial_{x_{II}},\partial_{x_{II}} \rangle = \tilde{\gamma} \langle A\partial_{\tau_I} - \partial_{x_I}
,A\partial_{\tau_I} - \partial_{x_I} \rangle = \tilde{\gamma}^2 (-A^2 + a_r^2), \nonumber \\
& \Rightarrow & \gamma = \tilde{\gamma} = \frac{1}{\sqrt{1-\frac{A^2}{a_r^2}}}.
\end{eqnarray}
This gives the coordinate relations
\begin{eqnarray}
\tau_{II} & = & \gamma \left( \tau_I + A x_I\right), \nonumber \\
x_{II} & = & -\gamma \left( \frac{A}{a_r^2} \tau_I + x_I\right), \Label{I-II-coord}\\
y_{II} & = & y_I, \quad \varphi_{II} = \varphi_{I}.
\end{eqnarray}
We can write the metric in region II in I-coordinates as
\begin{eqnarray}
ds_{II} & = & -d\tau_{II}^2 + a^2 (dx_{II}^2 + dy_{II}^2 + y_{II}^2 d\varphi_{II}^2) \\
& \approx & - d\tau_I^2 + a^2
 \left( dx_I^2 + dy_{I}^2 + y_{I}^2 d\varphi_{I}^2\right) + 2A\left(\frac{a^2}{a_r^2}
-1\right) d\tau_I \, dx_I, \Label{metric-II}
\end{eqnarray}
where we have dropped subleading terms suppressed by $1/a_r^2 \sim \exp(-2N)$ and $1/a^2\sim \exp(-2 N_* -2 N)$.

\paragraph{Boundaries of the regions:}
The boundaries of regions I, II and III are defined by null surfaces originating from the surface $(\tau=0,x=0,y,\varphi)$,
propagating in the $(\tau,x)$-plane. This implies $ds^2=0$ with $dy=d\varphi=0$, from which one can solve
\begin{equation}
\tau_I = \frac{(a_r x_I +1)^2-1}{2} \approx \frac{a_r^2x_I^2}{2} \Label{I-III-border}
\end{equation}
as the equation for the boundary between regions I and III; the approximation used was $a_r x \sim e^{N}e^{N_*-N} = e^{N_*} \ggg
1$. Since spacetime in region II is also RD-FRW with time coordinate $\tau_{II}$, the boundary between regions II and III is
given by $\tau_{II} \approx \frac{1}{2} a_r^2 x_{II}^2$.

\subsubsection{Region III} \label{sec-reg3}

 We now define the coordinates and the metric in region III. We do this by first constructing a physically motivated ansatz for
a coordinate system and interpolated metric. Then we vary this metric and interpolation, and compute certain coordinate invariant
quantities to show that our ansatz produces the physically most reasonable interpolation in region III.

We first fix the direction $\partial_{\tau_{III}}$ by using the symmetry of the spacetime by setting
\begin{equation}
\partial_{\tau_{III}} \propto \partial_{\tau_I} + \partial_{\tau_{II}} \propto \partial_{\tau_{I}} - \frac{A}{2a_r^2}
\partial_{x_I}.
\end{equation}
By orthogonality this fixes the coordinate $x_{III}$ to be of the form
\begin{equation}
dx_{III} \propto \frac{A}{2a_r^2} d\tau_I + dx_I, \quad \Rightarrow \quad x_{III} = h\left(\frac{A}{2a_r^2} \tau_I +
x_{I}\right), \Label{III-x}
\end{equation}
for some function $h$.

Since regions I and II are causally disconnected, identical RD-FRW spaces (with a relative boost between them), we expect evolution in each region to be identical. That is, we expect an event (e.g. recombination) that occurs in region I at $\tau_I = \t_0$ will occur in region II at $\t_{II} = \t_0$. We are therefore motivated to take an ansatz for region III that matches from a point $P=(\tau_{I}= \tau_0, x_I= \frac{\sqrt{2\tau_0}}{a_r})$ on the boundary between regions I and III to a point $\tilde{P}=(\tau_{II}=\tau_0,x_{II}=\frac{\sqrt{2\tau_0}}{a_r})$ on the boundary regions II and III along a surface of constant time ($\t_{III}$), see \figref{fig-as6}. Naively, inside region III this surface could take any shape, though clearly there are choices that are more physically motivated. The simplest physically motivated ansatz we can take is to demand the surfaces of equal time are linear in $(\tau_I,x_I)$ (and ($\tau_{II},x_{II})$)
coordinates. We will first make this choice, and then work out a more general ansatz in \secref{sec-gencoord}. We will show that to the order we are working in, all of these choices give similar results.

The slope from $P$ to $\tilde{P}$ is independent of $\t_0$ and this fixes  
\begin{equation}
\tau_{III} = f\left(\tau_{I} + \frac{A}{2} x_{I}\right), \Label{III-tau}
\end{equation}
for some function $f$. Finally, we can fix the functions $f$ and $h$ by choosing (without loss of generality) that the coordinates are continuous across the
border so that points $P$ and $\tilde{P}$ are given in III-coordinates as $P=(\tau_{III}=\tau_0,
x_{III}=\frac{\sqrt{2\tau_0}}{a_r})$ and $\tilde{P} = (\tau_{III}=\tau_0,x_{III} = -\frac{\sqrt{2\tau_0}}{a_r})$.\footnote{Note
that going from region II to region III, the $x$-coordinate changes sign; $x_{II}\to -x_{III}$. This is because $x_{II}$
increases towards the left, and $x_{III}$ increases to the right.} With these constraints the inverses of the functions are fixed
as
\begin{eqnarray}
f^{-1}(s) & = & s + \frac{A}{2a_r} \sqrt{2s}, \\
h^{-1}(s) & = & s + \textrm{sign}(s) \, \frac{A }{4} s^2,
\end{eqnarray}
where $s$ is a dummy variable, and the $\textrm{sign}(s)$ ensures that $h$ is an odd function in $s$. These can be
inverted to find $f$ and $h$, but for our purposes these inverses are sufficient.

Note that the coordinate system thus defined respects the symmetry of the spacetime; in II-coordinates we have
\begin{eqnarray}
\tau_{III} & = & f\left(\tau_{II} + \frac{A}{2} x_{II}\right), \\
x_{III} & = & -h\left(\frac{A}{2a_r^2} \tau_{II} + x_{II}\right).
\end{eqnarray}

Using the coordinate relations above and the RD-FRW metric \eqref{eq:RD}, we can find the metric in regions I and II in
III-coordinates. We compute these metrics at the boundaries\footnote{By `metric on the boundary' we simply mean the values of the
4D metric tensor $g_{\mu\nu}$ on the boundary, \emph{not} the induced 3D metric on the boundary.}
 of region III, and then \emph{define} the metric in region III as a
linear interpolation in $x_{III}$ on slices of constant $\tau_{III}$  between the boundaries.

Using the relations (\ref{III-x}) and (\ref{III-tau}), and the metric (\ref{eq:RD}), we can write the metric at point $P$ as
\begin{equation} 
ds_{III}^2|_{P} = - d\tau_{III}^2 + a^2 \left( 1+ \frac{A}{a_r}\sqrt{2\tau_{0}} \right) dx_{III}^2 + a^2 (dy_{III}^2 + y_{III}^2
d\varphi_{III}^2) - \frac{Aa^2}{a_r^2}  d\tau_{III} dx_{III}, \Label{met-bI}
\end{equation}
where here  $a = \sqrt{C \tau_{III}}$, and we have again neglected subdominant terms suppressed by $e^{-N}$, $e^{-N_*}$ and
$(A\epsilon)^2$. Similarly, we can find the metric at the opposite point $\tilde{P} $
\begin{equation} 
ds_{III}^2|_{\tilde{P}} = - d\tau_{III}^2 + a^2 \left( 1+ \frac{A}{a_r}\sqrt{2\tau_{0}} \right) dx_{III}^2 + a^2 (dy_{III}^2 +
y_{III}^2 d\varphi_{III}^2) + \frac{Aa^2}{a_r^2}  d\tau_{III} dx_{III}. \Label{met-bII}
\end{equation}
Since $P$ and $\tilde{P}$ were  arbitrary points on the boundaries, these relations specify the metric everywhere on the
boundaries of region III.

\paragraph{The interpolated metric:} We define the metric in region III as a linear interpolation in $x_{III}$ on the constant timeslice $\tau_{III}=\tau_0$ between \eqref{met-bI} and \eqref{met-bII}. Since $g^{III}_{\tau\tau}(P) = g^{III}_{\tau\tau}(\tilde{P})$ \footnote{While
we write $g^{III}_{\tau\tau}$ for brevity, we of course mean $g^{III}_{\tau_{III}\tau_{III}}$, and likewise for the other
components.}, the linear interpolation on the timeslice $\tau_{III}=\tau_0$ between $P$ and $\tilde{P}$ gives
\begin{equation}
g^{III}_{\tau\tau}(\tau_0,x_{III}) \equiv g^{III}_{\tau\tau}(P) = -1.
\end{equation}
Likewise, $g^{III}_{xx}, \, g^{III}_{yy}$ and $g^{III}_{\varphi\varphi}$ have the same value at $P$ and $\tilde{P}$, and hence we
fix those components to be independent of $x_{III}$ on the timeslice $\tau_{III}=\tau_0$.

The remaining component is $g^{III}_{\tau x}$, which changes sign between points $P$ and $\tilde{P}$. The linear interpolation
then defines $g^{III}_{\tau x}$  on the timeslice $\tau_{III}=\tau_0$ as
\begin{equation}
g^{III}_{\tau x}(\tau_0,x_{III}) \equiv g^{III}_{\tau x}(P) \cdot \left(\frac{x_{III}}{\sqrt{2\tau_0}/a_r} \right) = - \frac{Aa
x_{III}}{2},
\end{equation}
which reduces to the correct boundary values as $x_{III} \to \pm \frac{\sqrt{2\tau_0}}{a_r} = \pm \frac{a}{a_r^2}$.

We have completely defined the interpolated metric on the timeslice $\tau_{III}=\tau_0$, and since $\tau_0$ was arbitrary, this
defines the metric in all of region III as
\begin{equation}
ds^2_{III} =- d\tau_{III}^2 + a^2 \left( 1+ \frac{A}{a_r}\sqrt{2\tau_{III}} \right) dx_{III}^2 + a^2 (dy_{III}^2 + y_{III}^2
d\varphi_{III}^2) - Aa x_{III}  d\tau_{III} dx_{III}. \Label{gIII}
\end{equation}

\subsubsection{A general coordinate system} \label{sec-gencoord} 
The previous ansatz was the simplest physically motivated one. However, one could argue that a more general interpolation could capture the physics in region III better. In this section we will find a more general coordinate system, constrain it by physical considerations and then show that the effect on the galaxy flow is minimal between these choices. We vary our coordinate system by defining
\begin{equation}
\left\{ \begin{array}{lcl} T(\tau_{III},x_{III}) & = & \tau_{III} + A\delta_1(\tau_{III},x_{III}), \\
X(\tau_{III},x_{III}) & = & x_{III} + A\delta_2(\tau_{III},x_{III}), \end{array} \right. \Label{TX}
\end{equation}
where $\delta_1$ and $\delta_2$ are a priori arbitrary, but will be constrained by physical considerations. The factors of $A$
have been added for later convenience. Without loss of generality we pick $\delta_1$ and $\delta_2$ to vanish on the boundaries
of region III, to keep the coordinates continuous across the boundary. With this choice, it can be shown that the magnitudes of the $\delta$s are $|\delta_1| \lesssim \epsilon$ and $|\delta_2| \lesssim \epsilon^2$. If
$\delta_1$ and $\delta_2$ were bigger, $T$ and $X$ would not be timelike and spacelike respectively for generic $\delta_1$ and
$\delta_2$.\footnote{In the above we've assumed that $\delta_1$ and $\delta_2$ depend polynomially on the variables, so that
$\mathcal{O}(\partial_T {\delta_1}) \sim \mathcal{O}(\frac{\delta_1}{T})$ etc.  While more general forms could be considered,
they do not seem physically well motivated.}

Using (\ref{TX}) and (\ref{III-x},\ref{III-tau}) we solve for $\tau_{I}(T,X)$ and $x_I(T,X)$, and write the metrics in regions I
and II in $(T,X)$-coordinates. As above, we then interpolate linearly in $X$ across region III, giving the metric in region III
as
\begin{equation}
ds^2_{III} =- dT^2 + a^2(T) \left( 1+ \frac{A}{a_r}\sqrt{2T}(1-\delta_2^X) \right) dX^2 + a^2 (dy_{III}^2 + y_{III}^2
d\varphi_{III}^2) - Aa X (1-\delta_2^T) dT dX, \Label{gIII-TX}
\end{equation}
with $a(T) = \sqrt{CT}$, and we have defined
\begin{equation}
\left\{ \begin{array}{lcl} \delta_2^T & \equiv & -\textrm{sign}(X) 2a_r^2 \partial_T \delta_2(T,X), \\
\delta_2^X & \equiv & \frac{2a_r}{\sqrt{2T}}  \partial_X \delta_2(T,X) \, .\end{array} \right. 
\end{equation}
The appearance of $\textrm{sign}(X)$ above can be traced back to the the sign change $x_{II}\to -x_{III}$ across the boundary.
Due to the considerations above we generically have $\delta_2^T \sim \delta_2^X \sim 1$. Note that $\delta_1$ does not appear in
the metric to this order, and thus we are free to set it to vanish, giving $T = \tau_{III}$.

We must demand that the boundaries $T = \frac{1}{2} a_r^2 X^2$ between regions I and III (and II and III) are null in this metric. The boundary is generated by the vector $
a_r^2 X \partial_T + \partial_X$, and requiring this vector to be null gives
\begin{equation}
\delta_2^T(T,X) |_b = \delta_2^X(T,X)|_b, \label{delta-cons}
\end{equation}
where the $b$ indicates that this relation is valid on the boundary.

\paragraph{Curvature in region III:} At this stage, we have some constraints on the $\delta$s, but a lot of freedom remains. We can ask what physical expectations we have about the behavior of physical quantities in region III and use that to further constrain the functions. For example, if the $\delta$s were wildly varying functions, then the curvature in region III would be large in places, varying on small scales and possibly even negative. This behavior is clearly unphysical. We are thus motivated to examine the curvature scalar along a constant time surface to aid in constraining the $\delta$s. We will find that with this criteria as well as symmetry considerations there are two natural choices, and the difference in flow velocities between the two choices is at most $\mathcal{O}(20 \% -30 \%)$.

The curvature scalar  in region III is
\begin{equation}
R = \frac{2A}{a_r(2T)^{\frac{3}{2}}} \left( 3 + 3a_r^2 \partial_T \delta_2 + a_r \left( 3a_rX-4\sqrt{2T}\right)
\partial_T\partial_X \delta_2 + 2T a_r^2 \partial_T^2 \delta_2 + 2T a_r \left( a_rX-\sqrt{2T}\right) \partial_T^2 \partial_X \delta_2
\right).\Label{ricci}
\end{equation}
The constraints on $\delta_2$ to a large degree fix its functional form. By continuity of the coordinates $\delta_2$ must vanish as $X
\to \pm \frac{\sqrt{2T}}{a_r}$, and by the symmetry of the spacetime it must vanish as $X\to 0$. Furthermore, as we have
$\delta_2 \sim \epsilon^2$, a fairly general\footnote{More generally, a function of the form $\left(
\frac{a_r}{\sqrt{2T}}\right)^a X^b \left( \frac{2T}{a_r^2}-X^2 \right)^c$, with $2c+b-a=2$ and $b$ odd would work. Our choice,
$a=b=c=1$, seems physically the best motivated.} polynomial function  we can write down that satisfies \eqref{delta-cons} and the boundary conditions is
\begin{equation}
\delta_2(T,X) = F\left( \frac{a_r X}{\sqrt{2T}} \right) \, \frac{a_r X}{\sqrt{2T}} \left( X^2 - \frac{2T}{a_r^2} \right),
\Label{delta2}
\end{equation}
where $F$ is an unknown function of at most order unity, and $\frac{a_r X}{\sqrt{2T}} \sim \mathcal{O}(1)$.

\begin{figure}[t]
\begin{center}
\includegraphics[width=0.6\textwidth]{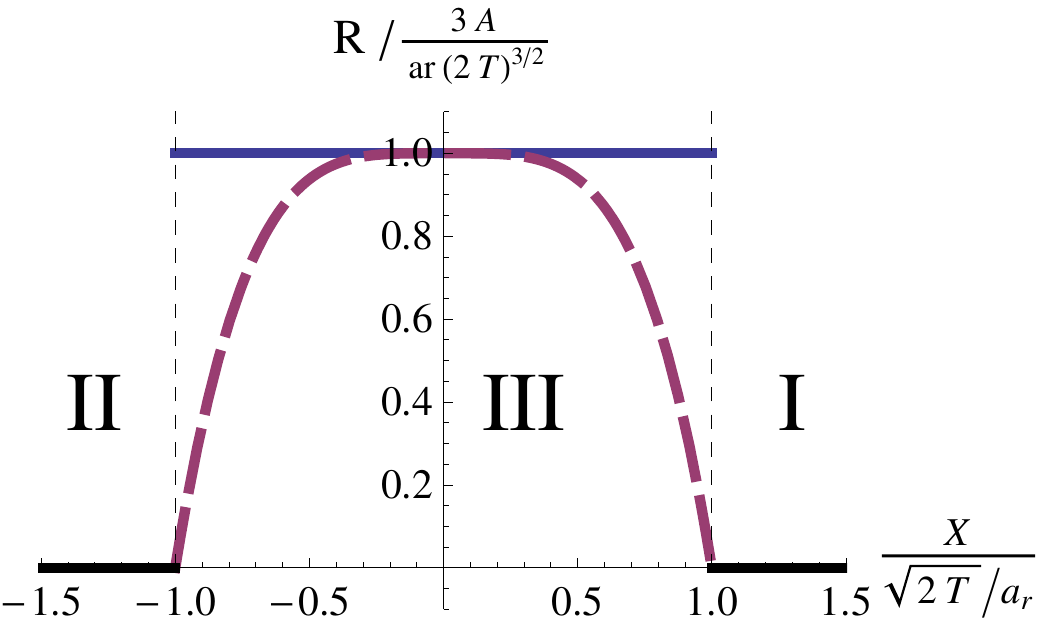}
\end{center}
\caption{\label{fig-ricci} Graph of the value of the Ricci scalar for the two proposed interpolated metrics. In regions I and II
($|X| > \frac{\sqrt{2T}}{a_r}$) the spacetime is RD-FRW and the Ricci scalar vanishes. For the $(\tau_{III},x_{III})$-metric the
scalar is a positive constant proportional to $\epsilon$ in region III given by the solid blue line, and for the $(b_0=-b_1=-\frac{3}{10})$ system discussed in the text the Ricci scalar is given by the dashed purple curve. }
\end{figure}

At the boundary between regions I and III we have $\frac{a_r X}{\sqrt{2T}}=1$, and hence we expand $F(s) = \sum b_n (1-s)^n$ for
some constants $b_n$. We are mostly interested in geodesics relatively close to the boundary, so to a good approximation we can
truncate
\begin{equation}
F\left( \frac{a_r X}{\sqrt{2T}} \right) \equiv b_0 + b_1\left(1- \frac{a_r X}{\sqrt{2T}} \right).
\end{equation}
While this is clearly not the most general form of $F$ in the interior of region III, it captures the physics close to the
boundary, and by a suitable choice of $b_0$ and $b_1$ we ensure that physics in the interior is also reasonable. We fix the
constants $b_0$ and $b_1$ by demanding that the curvature scalar vanishes on the boundary of the regions, and that the curvature scalar is differentiable at $X=0$, as there are no sources along $X=0$. Plugging
(\ref{delta2}) into (\ref{ricci}) we find that these constraints imply
\begin{equation}
b_0 = -b_1 = -\frac{3}{10}.
\end{equation}
Note that the choice $b_0=b_1=0$ gives the $(\tau_{III},x_{III})$ coordinate system and corresponds to a constant curvature across region III proportional to $\epsilon$. In \figref{fig-ricci} we plot the scalar
curvature on a constant time slice across the spacetime, for both $b_0=b_1=0$ and $b_0=-b_1=-\frac{3}{10}$.

In the next section we will show that the matter flow is largely unaffected between these two cases; the flow velocities in the $b_0
= -b_1 = -\frac{3}{10}$ are roughly 20\%-30\% higher than in the $b_0=b_1=0$ case. Thus our conclusions are fairly robust under the choice
of the interpolated metric.

\section{Galaxy motion}
In the previous section we found a physically motivated metric for region III. Together with the metrics in regions I and II and the various coordinate relations we have a global description of the spacetime background to order $\epsilon$. In this section, we will find the motion of galaxies through this spacetime and show that the collision generically leads to peculiar velocities in the affected region. 

In our approximation we will assume the matter that makes up galaxies was created at reheating, and was initially moving orthogonal to the reheating surface, i.e. initially generated by $\partial_{\tau_{I}}$ and $\partial_{\tau_{II}}$ in regions I and II respectively. We track the future evolution of these geodesics and analyze the motion of the galaxies as seen by present day observers.

\subsection{Timelike geodesics in region III}
We are interested in observers originating in region I, who have entered region III in relatively recent times\footnote{That is, we are most interested in observers who see a hot or cold spot that approximately takes up at most half of the CMB sky. A collision that takes up more than half of the CMB sky can be mapped in our approximation to one taking up less than half the sky by a boost. The former ``spot'' maps to the CMB rest frame and vice versa. To remain consistent within our approximations we will require that the CMB spot is less than about $80^\circ$, otherwise corrections of order $\epsilon^2$ can become important. It was argued in \cite{wwc2} and \cite{Freivogel:2009it} that collisions that take up the entire sky are not observable, at least in the CMB temperature map as they appear as a pure dipole only.}. We define the time of the observer to be $\t_{III} = \t_n \sim \mathcal{O} (e^{2N_*} )$. This implies the observer is at $x_{III} = x_n \sim \mathcal{O}(e^{N_*-N})$; for larger values of $x$ the
observers will not yet see the collision. We will find the timelike geodesics in region III for this range of parameters.

Galaxies that originate in region I  follow the geodesics with constant $x_{I},y_{I}$ and $\varphi_{I}$ while in region I. Upon entering region III they have four-velocity
\begin{eqnarray}
u_0 & = & \partial_{\tau_I} = u^{\tau}_0 \partial_{\tau_{III}} + u^{x}_0 \partial_{x_{III}}, \quad
\textrm{with} \Label{initial-u} \\
u^{\tau}_0 & = & 1 - \mathcal{O}(\epsilon^2), \quad u^{x}_0 = \frac{A}{2a_r^2}(1+\mathcal{O}(\epsilon^2)), 
\end{eqnarray}
where $u^{\tau}_0 \equiv u^{\tau_{III}}_0$ and $u^{x}_0 \equiv u^{x_{III}}_0$, and $u^{y}_0,\, u^{\varphi}_0 \approx 0
$.\footnote{A more careful computation yields $u^y \sim u^x \epsilon^2 \ll u^x$, and hence motion in the $y$-direction is
negligible in the order in which we work.} \footnote{If one does a similar computation for the CMB photons, they generically already have peculiar physical velocities of order 1 to satisfy the null condition. Therefore, the corrections due to the backreaction are subleading in $\epsilon$. If one computes their effect on the CMB temperature map as in \cite{wwc2} the effect from the backreaction multiplies a quantity that is already subleading, thus causing the effect to be $\mathcal{O}(\epsilon^2)$. It is therefore consistent at this order to ignore the backreaction effect for the CMB photons.}

We are thus interested in timelike geodesics in region III with initial velocity given above. Since the velocities are
non-relativistic\footnote{The velocities given are in comoving coordinates; the corresponding physical three-velocity is
$v\approx u^x a(\tau) \sim A \epsilon \ll 1$ for relevant times $\tau$, hence non-relativistic.}, the normalization condition
$|u|^2=-1$ gives
\begin{equation}
 u^{\tau} = 1 - \mathcal{O}(\epsilon^2), \quad \Leftrightarrow \tau = s \left(1+ \mathcal{O}(\epsilon^2) \right), \Label{u-tau}
\end{equation}
where $s$ is the affine parameter on the geodesic.  Making use of \eqref{u-tau}, we find the geodesic equation for $u^{x}$ to first order in $\epsilon$ to be
\begin{equation}
\frac{d^2x_{III}}{d\tau_{III}^2} + \frac{1}{\tau_{III}} \frac{dx_{III}}{d\tau_{III}} - \frac{Ax_{III}}{2a_r
(2\tau_{III})^{\frac{3}{2}}}
 = 0.
\end{equation}
This can be solved in the relevant range of coordinates to give 
\begin{equation}
x^{\textrm{geod.}}_{III}(\tau_{III}) = x_{0} +  \frac{A\tau_{0}}{a_r^2} \left[ \left(\sqrt{\frac{\tau_{III}}{\tau_0}}-1\right) -
\frac{1}{2} \left( 1-\frac{a_r^2 u^x_0}{2A} \right) \ln \frac{\tau_{III}}{\tau_0} \right] + \mathcal{O}(\epsilon^3),
\Label{mass-geod}
\end{equation}
where $(\tau_0,x_0)$ is the point where the geodesic intersects the boundary between regions I and III, as shown in figure
\ref{sketch}, and $u^x_0$ is the $x$-component of the four-velocity at that point.\footnote{For galaxies that originated in region II, the analysis is identical aside from a straightforward sign flip of the second term. A corresponding change is made for subsequent formulas. It turns out that our results for the observable quantities we compute later on automatically take this into account and so we suppress the further analysis of the region II galaxies here to simplify the presentation.  The physical results we compute later are completely general. \label{signfoot}} \, 

\paragraph{Peculiar velocity:} Using the initial velocity found in (\ref{initial-u}) and the solution for the geodesic
(\ref{mass-geod}), we find that at time $\tau_{III}$,  the velocity of a galaxy which entered region III at time $\tau_0$ is
given by
\begin{equation}
v_{\textrm{gal}} ^{III} = \frac{dx^{\textrm{geod.}}_{III}}{d\tau_{III}} = \frac{A}{2a_r^2} \sqrt{\frac{\tau_{0}}{\tau_{III}}}.
\Label{vgal1}
\end{equation}
Note that by velocity we mean velocity in coordinate system III.  To convert these to physical velocities, one must take into
account the velocity due to the expansion of the universe, i.e. Hubble's law and convert to the CMB rest frame.

\paragraph{The general coordinate system:} The procedure above can be repeated for the $(T,X)$-coordinate system. In this case,
the geodesic equation becomes
\begin{equation}
\frac{d^2X}{dT^2} + \frac{1}{T} \frac{dX}{dT} - \frac{AX}{2a_r (2T)^{\frac{3}{2}}} \left[ 1 + 4(b_0+b_1) \left(\frac{a_r
X}{\sqrt{2T}}\right)^3 - 12 b_1 \left(\frac{a_r X}{\sqrt{2T}}\right)^4 \right]
 = 0,
\end{equation}
and the initial velocity when the galaxy enters region III in these coordinates becomes
\begin{equation}
u^X_0 = \frac{A}{2a_r^2} \left( 1 - 4 b_0\right).
\end{equation}
The geodesic equation can be solved perturbatively to first non-trivial order in $\epsilon$ to give
\begin{equation}
X^{\textrm{geod.}}(T) = X_0 + \frac{AT_0}{a_r^2} \left[ \left( \sqrt{\frac{T}{T_0}} -1\right) - (b_0-\frac{b_1}{3}) +
(b_0+b_1)\left( \frac{T_0}{T} + \ln \frac{T_0}{T} \right) - \frac{4}{3} b_1 \left( \frac{T_0}{T} \right)^{\frac{3}{2}}\right],
\end{equation}
which reduces to (\ref{mass-geod}) as $b_0,b_1 \to 0$. Here $(X_0,T_0)$ is the point where the galaxy enters region III. Differentiating, we find the velocity along the geodesic as
\begin{equation}
v_{\textrm{gal}}^{(T,X)} = \frac{dX^{\textrm{geod.}}}{dT} = \frac{A}{2a_r^2} \left[ \sqrt{\frac{T_0}{T}} - 2(b_0+b_1)
\left(\frac{T_0}{T} + \frac{T_0^2}{T^2} \right) + 4b_1 \frac{T_0^4}{T^4} \right].
\end{equation}

\begin{figure}[t]
\begin{center}
\includegraphics[width=0.6\textwidth]{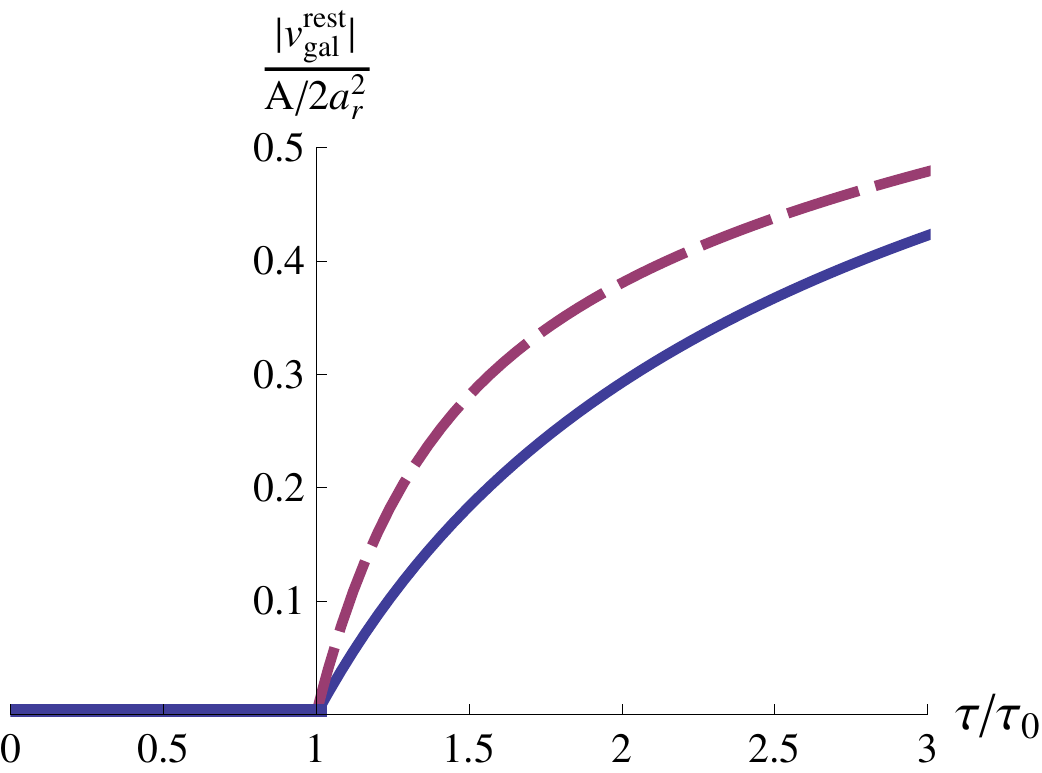}
\end{center}
\caption{\label{fig-vels} Effect of the choice of coordinates on the flow velocities in appropriately normalized units, in the rest frame of the CMB (see text for discussion of the rest frame). The
$\{b_0=b_1=0\}$ solution is portrayed with a thick blue line, while the $\{b_0=-b_1=-\frac{3}{10}\}$ case is shown with a dashed purple line. For $\t<\t_0$ the galaxies are in region I and hence have no flow velocity.}
\end{figure}

To assess the effect of the choice of the interpolating coordinate system on the matter flow, we convert the velocity above to the (approximate) CMB rest frame\footnote{We are focusing on collision scenarios were the affected CMB region takes up less than half the sky (again if it takes up more, we can map it to a scenario where it takes up less than half by a boost), i.e. the observer's CMB sky is dominated by photons that originate in region I. Because of this, the CMB rest frame is that of region I to a good approximation.} which is given by the coordinates of region I using the coordinate transformations (\ref{TX}), (\ref{III-tau}) and (\ref{III-x}). This
gives
\begin{equation} \label{vgal}
v_{\textrm{gal}}^\trm{rest} = -\frac{A}{2a_r^2} \left(1-\sqrt{\frac{\tau_0}{\tau_I}} \right) \left( 1 - 2(b_0+b_1) \left[
\sqrt{\frac{\tau_0}{\tau_I}} + \left( \frac{\tau_0}{\tau_I}\right)^{\frac{3}{2}} \right] + 4b_1 \left( \frac{\tau_0}{\tau_I}
\right)^2 \right),
\end{equation}
and in \figref{fig-vels} we plot these velocities for the two preferred sets of values, $b_0=b_1=0$ and
$b_0=-b_1=-\frac{3}{10}$. Both sets lead to comparable flow velocities, with the $b_0=-b_1=-\frac{3}{10}$
case generally leading to velocities that are roughly 20-30\% higher than the $b_0=b_1=0$ case.  This shows that our model is
fairly robust under the choice of a (physical) interpolating metric. For this reason, and for computational simplicity, we focus
on the $b_0=b_1=0$ case from now on; the results obtained for the other choice of interpolation will be similar,
but would require numerical analysis.

\subsection{Past lightcone of observers in region III}
In the previous section we found a general expression for the peculiar velocity of galaxies relative to the CMB rest frame \eqref{vgal}. We would like to find the observable consequences of these peculiar velocities, i.e. how big a patch of the sky they take up at a given redshift and how large the velocities are for reasonable scenarios. In this section we will work out how the size of the disk on the sky that contains galaxies with non-vanishing peculiar velocity varies as a function of $N$, $N_*$, the redshift to the galaxies in question and also in terms of the effect on the CMB sky from the collision found in \cite{wwc2}. To do this, we take an observer who has recently entered region III, project back his past lightcone to a given redshift and then find the galaxies behavior within this lightcone. We will focus on redshifts of approximately $0\lesssim z \lesssim1$.

\label{sec:lightcone} Consider observers at point $(\tau_n,x_n,y_n,\varphi_n)$ in region III (and region III coordinate values), travelling along one of the  geodesics
found above \eqref{mass-geod}. We wish to analyze the motion of galaxies as seen by these observers.  As neither the metric nor the scalar field depend on $y$ up to order $\mathcal{O}(\epsilon^2)$, we are free to
choose the origin of the $(y,\varphi)$ plane to coincide with the location of the observers, so that $y_n=0$. To find the past
light cone, we define a coordinate system centered on the observers, $(\tau_{\textrm{obs}},\xi_{\textrm{obs}},y_{\textrm{obs}},\varphi_{\textrm{obs}})$, related to the usual region III coordinates by
\begin{eqnarray}
\tau_{\textrm{obs}}=\tau_{III}, \quad & & \quad x_{\textrm{obs}} = x_{III} - x_n, \\
y_{\textrm{obs}} = y_{III}, \quad & & \quad \varphi_{\textrm{obs}}=\varphi_{III}.
\end{eqnarray}

\begin{figure}[t]
\begin{center}
\includegraphics[width=0.6\textwidth,angle=0]{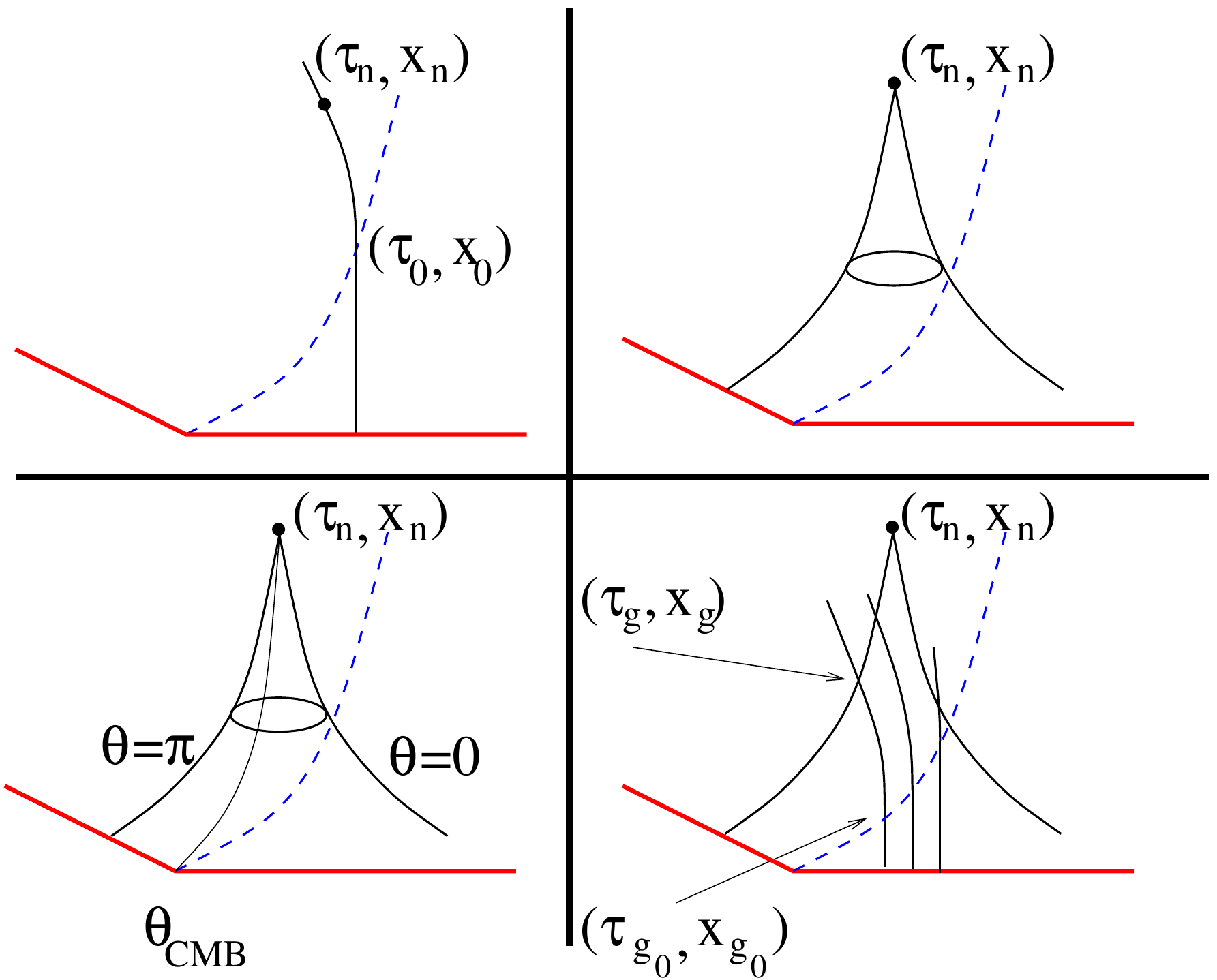}
\end{center}
\caption{\label{sketch} Sketch of the motion of the observers, their past lightcone, and other galaxies inside that lightcone.
The figure is not drawn to scale.}
\end{figure}

We can find the null geodesics projected back from the observer in a similar manner to the procedure for finding the timelike geodesics of the previous sections
\begin{equation}
\tau_{\textrm{obs}}(\xi_{\textrm{obs}}) = \frac{1}{2} \left( \sqrt{2\tau_n} - a_r\xi_{\textrm{obs}} \right)^2
\left(1+\mathcal{O}(A\epsilon) \right) \Label{lightcone-sphe},
\end{equation}
where $\xi_{\textrm{obs}} = \sqrt{x_{\textrm{obs}}^2 + y_{\textrm{obs}}^2}$. Note that the leading term gives the parabolic past
lightcone of RD-FRW space. Using the leading term of the past lightcone is sufficient for our purposes, as the velocity
(\ref{vgal1}) is already linear in $\epsilon$, and thus any subleading corrections to the lightcone will lead to sub-subleading corrections
to the velocities of the galaxies, and will therefore be negligible to our purposes. The lightcone is illustrated in
\figref{sketch}.

The lightcone can be written in a more convenient form. Consider  observers looking in a fixed direction $\theta_{\textrm{obs}}$ in the night sky, independent of $\varphi_{\textrm{obs}}$. This translates to the III coordinates as $\tan \theta_{\textrm{obs}} =
\frac{y_{\textrm{obs}}}{x_{\textrm{obs}}} = \frac{y_{III}}{x_{III}-x_n} =$ constant, which implies $y_{III} = \tan
\theta_{\textrm{obs}} (x_{III}-x_n)$. Using this, the lightcone (\ref{lightcone-sphe}) can be written as
\begin{equation}
x_{III}-x_n =  \left(\frac{\sqrt{2\tau_n}}{a_r} - \frac{\sqrt{2\tau_{III}}}{a_r}\right)  \cos \theta_{\textrm{obs}}.
\Label{lightcone}
\end{equation}

\paragraph{The collision disk:} In \cite{wwc2} it was shown that a bubble collision generically produces a hot or cold spot on the sky. It is useful to ask how the galaxy flow will depend on the size and temperature profile of this spot. To that end, we express our results in terms of the angular radius of the affected disk on the CMB sky, given by $\pi-\theta_{\textrm{CMB}}$.\footnote{Recall that in our coordinates $\theta=\pi$ points towards the collision} The intersection of the two sloped parts of the reheating surface is given by $\tau_{III}=x_{III}=0$. Plugging this into the lightcone
(\ref{lightcone}) gives the size of the collision disk seen by the observers as (see \figref{sketch})
\begin{equation}
\cos \theta_{\textrm{CMB}} = -\frac{a_r x_n}{\sqrt{2\tau_n}}.
\end{equation}

\paragraph{Observable galaxies:}  Consider other galaxies as seen by our observers; illustrated in  \figref{sketch}.
The intersection of the geodesic of the other galaxy with the past lightcone of our observers is given by the equations
\begin{eqnarray}
\textrm{Galaxy geodesic:} \quad x_{g} - x_{g_0} & = & \frac{A\tau_{g_0}}{a_r^2} \left( \sqrt{\frac{\tau_{g}}{\tau_{g_0}}} -1\right) ,\\
\textrm{Lightcone:} \quad x_g - x_n & = & \left( {\sqrt{2 \tau_n} \over a_r}-{\sqrt{2 \tau_g} \over a_r} \right) \cos \theta_{\textrm{obs}} ,
\end{eqnarray}
where $(\tau_g,x_g)$ is the point of intersection and $(\tau_{g_0},x_{g_0})$ is the point where the other galaxy crossed the boundary
between regions I and III; refer again to \figref{sketch}.

The peculiar velocity of the galaxy (in region III coordinates) at the intersection is given by (\ref{vgal1}) as $v^{III}_{gal} = \frac{A}{2a_r^2}
\sqrt{\tau_{g_0}/\tau_g}$. To write this in variables more appropriate to the observer, we first use the equation for the geodesic of the galaxy
to solve for $\tau_{g_0}=\tau_{g_0}(\tau_g,x_g)$. As $(\tau_{g_0},x_{g_0},y_{g_0})$ is on the boundary between the regions, we have $x_{g_0} =
\frac{\sqrt{2\tau_{g_0}}}{a_r}$, and we get
\begin{equation}
\frac{\sqrt{2\tau_{g_0}}}{a_r} = x_g + \frac{Ax_g}{2}\left( x_g - \frac{\sqrt{2\tau_g}}{a_r}\right).
\end{equation}
We insert this into the velocity, and eliminate $x_g$ in favour of $x_n$, $\theta_{\textrm{obs}}$ and $\tau_n$ using the equation
for the lightcone. Furthermore, as light in FRW space redshifts as $\propto 1/a$, the redshift $z$ is given by  $1+z \equiv
\sqrt{\frac{\tau_n}{\tau_g}}$.  The velocity to leading order is
\begin{equation}
v_{\textrm{gal}} ^{III} \approx \frac{A}{2a_r^2} \left[ \frac{a_r x_n}{\sqrt{2\tau_n}} (1+z) + z \cos \theta_{\textrm{obs}}\right] =
\frac{A}{2a_r^2} \left[ -\cos \theta_{\textrm{CMB} } (1+z) + z \cos \theta_{\textrm{obs}}\right], \Label{vgal3}
\end{equation}
as a function of the direction in the night sky that is being observed, the angular radius of the CMB collision disk, and the redshift to the observed galaxy.\footnote{This formula is general and therefore also valid for galaxies that originate in region II, taking into account the sign flip pointed out in footnote \ref{signfoot}.}

\paragraph{The CMB frame:} The velocity computed above is in III-coordinates; it can be transformed back to I-coordinates, making up the (approximate)
rest frame of the CMB, using the coordinate relations (\ref{III-tau}) and (\ref{III-x}). This gives the velocity relative to the CMB rest frame as
\begin{equation}
v_{\textrm{gal}}^{\textrm{rest}} = v_{\textrm{gal}} ^{III} - \frac{A}{2a_r^2}\left(1+\mathcal{O}(\epsilon)\right)  \approx -\frac{A}{2a_r^2} \left[ 1+\cos
\theta_{\textrm{CMB}} (1+z) - z \cos \theta_{\textrm{obs}}\right]. \Label{vgal4}
\end{equation}
The relation above is valid in region III; galaxies in region I are moving along $\partial_{\tau_I}$ and thus have no velocity
with respect to the CMB frame. We can use this to find the size of the disk of affected galaxies by setting the velocity
$v_\trm{gal}^\trm{rest}$ to vanish for a fixed redshift $z$, in which case (\ref{vgal4}) implies
\begin{equation}
\cos \theta_{\textrm{gal}} (z)= \frac{1}{z} \left( 1+\cos \theta_{\textrm{CMB}} (1+z)\right) = \cos
\theta_{\textrm{CMB}} + \frac{1+\cos \theta_{\textrm{CMB}}}{z}. \Label{affected}
\end{equation}
All the galaxies inside this disk, $\theta_{\textrm{obs}} > \theta_{\textrm{gal}}$, are inside region
III and thus are affected by the collision. Note that $ \theta_{\textrm{gal}} < 
\theta_{\textrm{CMB}}$\footnote{Recall again, the angular radius of each disk is given by $\pi-\theta$ respectively.}, and hence the disk of affected galaxies is larger than the collision disk, which is also
clear from \figref{sketch}. Inside the disk of affected galaxies, we can see that the velocity at a fixed redshift decreases linearly with $\cos \theta_{\textrm{obs}}$ from its maximum value to zero at the edge of the disk.

\section{Results}

In this section we examine the effect of a bubble collision on the large scale galaxy flow in two scenarios:
\begin{enumerate}
\item{ A collision that produces a small spot on the CMB sky, akin to the WMAP cold spot.}
\item{A collision that produces a larger disk on the sky.}
\end{enumerate}
We will find that making our results agree with CMB observations will constrain our possible flows.

\paragraph{Physical velocities:} In the previous section we found the comoving peculiar velocities of galaxies inside the observer's past lightcone \eqref{vgal4}. To convert them to a physical velocity that can be observed we multiply by the scale factor at the time the galaxy intersected with the observer's past lightcone to get (we work in units where the speed of light, $c=1$, and our expression is valid to $\mathcal{O}(\epsilon)$)%
\begin{eqnarray}
v_{\textrm{phys}}  & = & v_{\textrm{gal}}^{\textrm{rest}} a(\tau_g) = -\frac{A\, a(\tau_n)}{2a_r^2}  \left(\frac{ 1+\cos
\theta_{\textrm{CMB}} + z (\cos \theta_{\textrm{CMB}} - \cos
\theta_{\textrm{obs}})}{1+z}\right) \nonumber \\
& = & -  \left(\frac{ 1+\cos \theta_{\textrm{CMB}} + z (\cos \theta_{\textrm{CMB}} - \cos
\theta_{\textrm{obs}})}{1+z}\right) \, Ae^{N_*-N}, \Label{vgal5}
\end{eqnarray}
where we wrote $a(\tau_g)$ in terms of the redshift and $a(\tau_n)$, and in the last line we used $\frac{a(\tau_n)}{a_r}\frac{1}{a_r} =
2e^{N_*-N}$.

\paragraph{Effect on the CMB:} It will be natural to express the galaxy flows we find in terms of the corresponding effect from the collision on the CMB itself. This effect was found in \cite{wwc2} and we refer the reader there for details. The redshift seen by an observer is defined
as the ratio of the energies of the photon as it was emitted and absorbed,
\begin{equation}
r_s \equiv \frac{E_n}{E_r} = \frac{a(\tau_r)}{a(\tau_n)}={T_n \over T_r}.
\end{equation}
For the unperturbed part of the reheating surface this is simply given by $r_s =a_r/a_n  = e^{-N_*} = T_n ^0 / T_r$.  In the
direction of the collision, $ \theta_{\textrm{obs}} > \theta_{\textrm{CMB}}$, reheating happens slightly  later (earlier) if  $A>0$ ($A<0$) (see \figref{fig-as6}) and the
redshift is given by 
\beq
r_s &=& \sqrt{\frac{\frac{1}{2} - A x_r}{\tau_n}} = e^{-N_*} \sqrt{1 - 2A x_r} \approx e^{-N_*} \left( 1 -
\frac{A\sqrt{2\tau_n}}{a_r} \left[ \cos \theta_{\textrm{obs}} - \cos \theta_{\textrm{CMB}} \right] \right) \nonumber \\ &\approx& {T_n ^0 \over T_r} \left(1 + {\Delta T \over T_n ^0} \right) .
\eeq
Here $x_r<0$ is the $x$-coordinate of the intersection of the observer's past lightcone with the perturbed reheating
surface, in a given direction $\theta_{\textrm{obs}}$, and we have expressed $x_r$ in terms of more physical quantities using the
equation for the lightcone, (\ref{lightcone}).  The variation from the average temperature due to the collision is\footnote{There will of course be the usual Gaussian fluctuations in the temperature map. They are not important for our analysis here, but a full treatment taking them into account is given in \cite{wwc2}.} 
\begin{equation}
\frac{\Delta T}{T_n^0} = \frac{A\sqrt{2\tau_n}}{a_r}  \left(\cos \theta_{\textrm{CMB}} - \cos
\theta_{\textrm{obs}}\right) = \left(\cos \theta_{\textrm{CMB}} - \cos \theta_{\textrm{obs}}\right) \left( 2A
e^{N_*-N}\right) , \Label{exper}
\end{equation}
and as $\Delta T/T_n^0$ is restricted by experiments, this expression can be used to give the matter flow velocities in terms
of experimental data. 

The largest flows and temperature variations are always in the direction of the collision, $(
\theta_{\textrm{obs}}= \pi)$, in which case we get
\begin{equation}
A e^{N_*-N} \approx \frac{1}{2} \frac{1}{1+\cos \theta_{\textrm{CMB}}} \left( \frac{\Delta T}{T^0_n} \right)_{\textrm{max}}.
\end{equation}
We can use this to express the flow velocity in terms of the temperature of the spot%
\beq%
v_\trm{phys} \approx - \frac{1}{2} \left( {\Delta T \over T_n^0} \right)_\trm{max} {1 \over 1+z} \left[ 1 + z \left( {\cos \theta_\trm{CMB}-\cos \theta_\trm{obs}  \over 1+\cos \theta_\trm{CMB} } \right) \right] . \label{spotflow}
\eeq% 
The largest flow we can measure is\footnote{Of course, there will be large fluctuations based on random local behavior for a given galaxy cluster. What we mean here is the maximum coherent contribution from a collision in our scenario.}
\begin{equation}
\left| v_{\textrm{phys}}^{\textrm{max}} \right| \sim \frac{1}{2}  \left| \frac{\Delta T}{T^0_n} \right|_{\textrm{max}}; \Label{vmax}
\end{equation}
note that this is independent of the size of the collision disk and the redshift.

\subsection{Scenario 1 : A small collision disk}
\begin{figure}[t]
\begin{center}
\includegraphics[width=0.6\textwidth,angle=0]{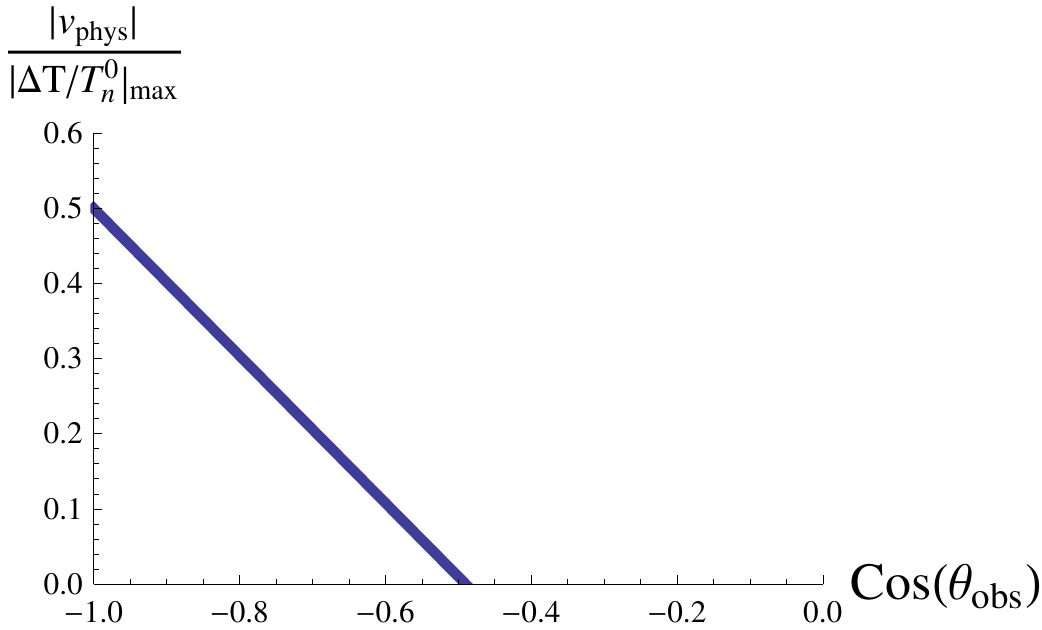}
\end{center}
\caption{\label{vsmallprof} Flow velocity for the small collision disk of scenario 1 at redshift $z=.02$ as a function of the angle on the sky. The velocity profile for general redshifts looks similar, always decreasing from the same maximum value to zero linearly in $\cos \theta_\trm{obs}$ to the edge of the disk. Outside the disk the peculiar velocity always vanishes.}
\end{figure}

We first examine the case of a small collision disk. If we take the WMAP cold spot as a rough guide, it is $\sim 73 \mu K$ colder on about $5^\circ$ scales \cite{coldspot1,coldspot2,coldspot3,coldspot4,coldspot5}. It was shown in \cite{wwc2} that the effect on the CMB of a collision is generically a spot, with a temperature profile that is linearly decreasing in $\cos \theta_{\textrm{obs}}$ from the center of spot to its edge, hence its average temperature is $\approx T_{\textrm{spot}}/2$, where $T_{\trm{spot}}$ is the temperature at the center of the spot. To give us an idea of the type of flows a collision that causes a spot like this generates, we take $T_n^0-T_\trm{spot} = 300 \mu K$ and angular radius $180^\circ -\theta_\trm{CMB} \approx 7^\circ$. This gives a flow velocity profile given by \eqref{spotflow} with $\left( \Delta T / T_n^0 \right)_\trm{max} \approx 5.5 \times 10^{-5}$. The maximum detectable flow velocity is
\beq
|v_\trm{phys} ^\trm{max} | \sim \mathcal{O}(5\cdot10^{-5} ) \sim \mathcal{O} (15 \, \trm{km/s}) .
\eeq

\begin{figure}[t]
\begin{center}
\includegraphics[width=0.6\textwidth,angle=0]{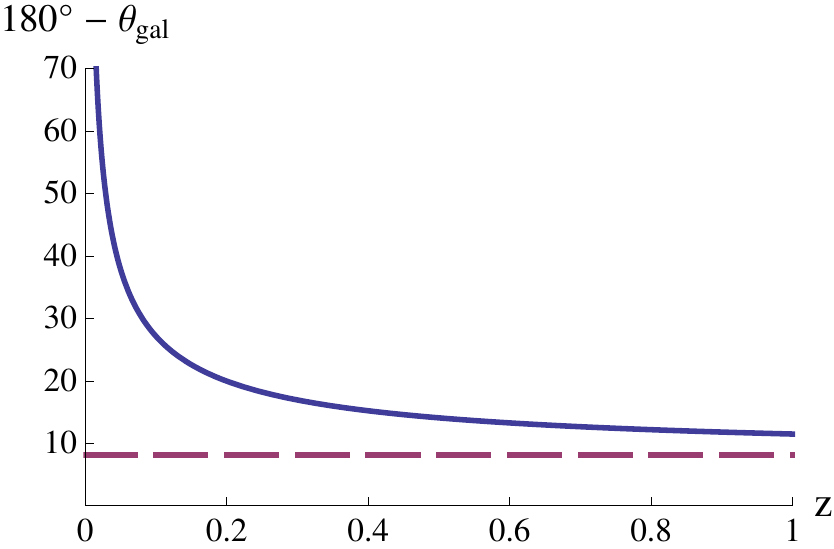}
\end{center}
\caption{\label{smalldisksize} Angular radius of the disk of affected galaxies for the small collision disk example of scenario 1. The solid blue line is the size of the galaxy disk and the dashed purple line gives the (constant) size of the cold spot for comparison. Note the galaxy disk can be significantly larger than that of the CMB spot, especially at low redshift.}
\end{figure}

At a given redshift $z$, the magnitude of the velocity inside the disk of affected galaxies will decrease from the redshift independent maximum value $|v_\trm{phys} ^\trm{max}|$ to zero linearly in $\cos \theta_\trm{obs}$ at the edge of the disk, we display a typical profile in \figref{vsmallprof}. The size of the disk is given in \eqref{affected}. In \figref{smalldisksize} we plot the angular radius of this disk as a function of redshift. We see that the angular size of the galaxy disk is always larger than that of the corresponding CMB spot, particularly at lower redshifts.

\subsection{Scenario 2 : A large collision disk}
\begin{figure}[t]
\begin{center}
\includegraphics[width=0.6\textwidth,angle=0]{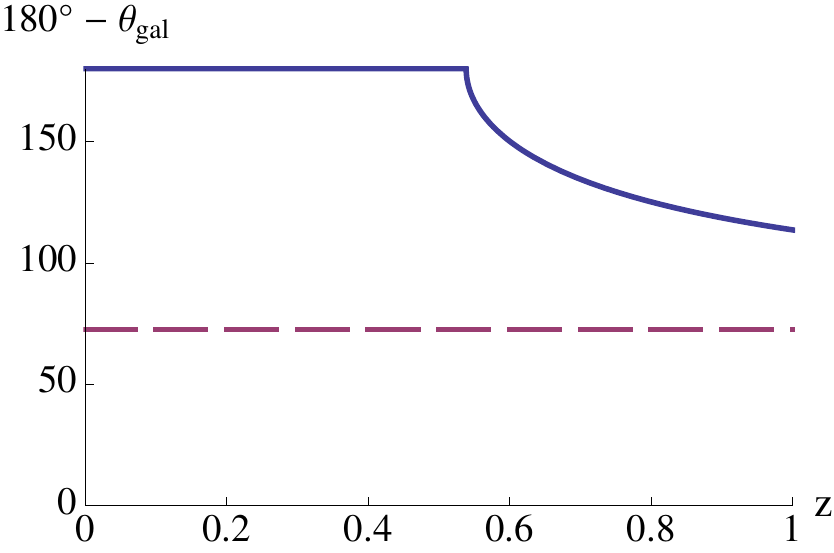}
\end{center}
\caption{\label{largedisksize} Angular radius of the disk of affected galaxies for the large collision disk example of scenario 2. The solid blue line is the size of the galaxy disk and the dashed purple line gives the (constant) size of the CMB spot for comparison. Note the galaxy disk can be significantly larger than that of the CMB spot. At lower redshifts we observe a flow across the entire sky.}
\end{figure}

Here we examine a larger collision disk, but with subsequently smaller $\Delta T$ so that it can still agree with present observational constraints. Such a disk could still contribute in the CMB data, for such effects as the hemispherical power asymmetry \cite{Eriksen:2007pc} or spatial correlation of the lower multipoles. We take $\cos \theta_\trm{CMB} = -0.3$, corresponding to a disk of angular radius $\approx 70^\circ$ on the sky and a corresponding $(\Delta T / T_n ^0)_\trm{max} = 5 \times 10^{-5}$. We then find a disk of affected galaxies of angular size in \figref{largedisksize} and maximum flow velocity
\be
|v_\trm{phys} ^\trm{max}| \sim \mathcal{O} (10 \, \trm{km/s} ) .
\ee

In \figref{largedisksize} we see that for redshifts out to $z \approx 0.6$ we observe a coherent flow across the entire sky. Even out to a redshift of one we still see a significant fraction of the sky flowing.

\section{Conclusions and comparison with observation}

In this paper we have shown that bubble collisions generically lead to a large scale bulk flow of galaxies. There is a disk on the sky of affected galaxies, always larger than the corresponding spot for the CMB temperature map, that decreases in size with redshift (though it may take up the entire sky for all redshifts observed). Within this disk, galaxies will coherently flow, with a peculiar velocity that decreases in magnitude from the center of the disk and is proportional to the corresponding temperature at the center of the spot in the CMB map. This is a distinctive signal profile and is in principle measurable with improved observations or analysis of current cluster catalogs. It would be very interesting if a bulk galaxy flow could be correlated with a corresponding disk or spot in the CMB data as this would be evidence for bubble collisions and the string landscape.

Recently, two groups have claimed to observe bulk flows of galaxy clusters \cite{Watkins:2008hf,darkflow,Kashlinsky:2008us}. They both measure a flow of $\mathcal{O}(400-800 \, \trm{km/s})$ out to a redshift $z \sim 0.2$. Both measurements are subject to a variety of random and systematic uncertainties (e.g. filtering methods and the conversion from temperature to velocity). They seem to have reasonable directional agreement, but the flow measured by Watkins et. al is considerably smaller than that of Kashlinsky et al. and they measure over different redshift ranges (Watkins et. al catalog is dominated by objects with $z<0.03$ and Kashlinsky by objects with $0.04<z<0.2$). The flows we find appear to be an order of magnitude smaller, as our flows are constrained by the corresponding effect in the CMB. Larger flows would lead to a very cold or hot spot in the CMB, likely ruled out by current observations. In addition, since our peculiar velocity profile falls off the further we are observing from the center of the disk of affected galaxies, and the disk itself may not take up the whole sky, the corresponding average peculiar velocity measured would be smaller. However, we have made several approximations which could affect the measured flow:
\begin{itemize}
\item{We have ignored any self-interaction and gravitation of the radiation and matter fluids. It is possible that as the fluids clump, this could affect the backreacted metric in region III and hence the flow.}
\item{We have ignored the transition to matter domination. Taking this transition into account would likely increase the flow for a subtle reason. In a matter dominated universe peculiar velocities will redshift more quickly. This implies that galaxies entering region III will transition more quickly to flowing along $\partial_{\tau_{III}}$. Transforming back to the CMB rest frame this results in higher {\it measured} peculiar velocity with respect to the rest frame.}
\item{The small freedom we have in choosing the metric interpolation in region III can have an $\mathcal{O}(1)$ effect on the flow. The flows computed above were for the choice that seems to produce the smallest flow. Other choices increase the flow velocity.}
\item{We have likely not accounted for many more detailed effects of the cosmological evolution that could affect the flow.}
\end{itemize}
Because of these approximations, it is not hard to envision a more detailed treatment finding an enhanced flow.

Many open questions on bubble collisions remain. It would be interesting to carry out a more rigorous model of cosmological evolution and perturbation theory in this background and find a more accurate picture of these effects. In this paper and \cite{wwc,wwc2} we have mapped out two possible effects of bubble collisions resulting from the string landscape, namely the effect on the CMB temperature map and on galaxy cluster motion. One could search for other effects, e.g. polarization of the CMB as our toy models bear some relation to those of \cite{Dvorkin:2007jp} and it may be possible to further correlate these signals. This effect could be particularly interesting in light of new polarization data expected from the Planck experiment. One could also look for signals in 21cm radiation data. On the observational side, most of the current measurements of the CMB have been presented in momentum ($\ell$) space, yet these effects are most apparent in real space. Each collision would lead to a disk on the sky and one could search the CMB and other data for traces of disks, some of which might overlap. While many of these things are likely to be difficult or require improved measurements (e.g. 21cm telescopes), we feel that cosmology and astrophysics offers a unique opportunity to possibly detect effects from high energy physics and string theory.

\section*{Acknowledgements}
We thank Spencer Chang, Lam Hui, Eugene Lim and Kris Sigurdson for helpful discussions. We are especially grateful to Matthew Kleban for collaboration during the early stages of this work. KL and TSL were supported in part by the Natural Sciences and Engineering Research Council of Canada and the Institute of Particle Physics.

\bibliographystyle{utphys}

\bibliography{bubble}

\end{document}